\documentclass[final,3p,times,twocolumn]{elsarticle}

\usepackage{graphics}

\usepackage{amssymb}
\usepackage{amsmath}
\usepackage{textcomp}
\usepackage{url}
\usepackage{tabularx}

\biboptions{authoryear}

\journal{JQSRT}

\newcommand{\ee}[1]{$\cdot$10$^{#1}$}
\newcommand{\degree}{\ensuremath{^\circ}}

\newcommand{\meanradI}{\ensuremath{I_{\rm mean,I}}}
\newcommand{\meanradQ}{\ensuremath{I_{\rm mean,Q}}}
\newcommand{\meanradU}{\ensuremath{I_{\rm mean,U}}}
\newcommand{\meanradV}{\ensuremath{I_{\rm mean,V}}}
\newcommand{\rms}{\ensuremath{\Delta_{\rm RMS}}}
\newcommand{\rmsI}{\ensuremath{\Delta_{\rm RMS,I}}}
\newcommand{\rmsQ}{\ensuremath{\Delta_{\rm RMS,Q}}}
\newcommand{\rmsU}{\ensuremath{\Delta_{\rm RMS,U}}}
\newcommand{\rmsV}{\ensuremath{\Delta_{\rm RMS,V}}}
\newcommand{\relstd}{\ensuremath{\sigma_{\rm rel}}}
\newcommand{\relstdI}{\ensuremath{\sigma_{\rm rel,I}}}
\newcommand{\relstdQ}{\ensuremath{\sigma_{\rm rel,Q}}}
\newcommand{\relstdU}{\ensuremath{\sigma_{\rm rel,U}}}
\newcommand{\relstdV}{\ensuremath{\sigma_{\rm rel,V}}}

\begin{document}

\begin{frontmatter}



\title{IPRT polarized radiative transfer model intercomparison project
  -- three-dimensional test cases (phase B)}
\author[lmu]{Claudia Emde}
\ead{claudia.emde@lmu.de}
\author[tro]{Vasileios Barlakas}
\author[lil]{C\'eline Cornet}
\author[col]{Frank Evans}
\author[nui]{Zhen Wang}
\author[lil]{Laurent C.-Labonotte}
\author[tro]{Andreas Macke}
\author[lmu]{Bernhard Mayer}
\author[lei]{Manfred Wendisch}

\address[lmu]{Meteorological Institute, Ludwig-Maximilians-University,
  Theresienstr. 37, Munich, Germany}
\address[tro]{Leibniz Institute for Tropospheric Research, Permoserstr. 15, Leipzig, Germany}

\address[lil]{Laboratoire d'Optique Atmosph\'erique, Universit\'e Lille, France}
\address[col]{University of Colorado, Boulder, CO 80309, USA}
\address[nui]{Nanjing University of Information Science and Technology,
  China}

\address[lei]{Leipzig Institute for Meteorology, University of Leipzig, Stephanstr. 3, Leipzig, Germany}

\begin{abstract}
Initially unpolarized solar radiation becomes polarized by scattering
in the Earth's atmosphere. In particular molecular scattering
(Rayleigh scattering) polarizes electromagnetic radiation, but
also scattering of radiation at aerosols, cloud droplets (Mie scattering) 
and ice crystals polarizes. Each atmospheric constituent produces a
characteristic polarization signal, thus spectro-polarimetric
measurements are frequently employed for remote sensing of aerosol and
cloud properties. 

Retrieval algorithms require efficient radiative transfer
models. Usually, these apply the plane-parallel approximation (PPA),
assuming that the atmosphere consists of
horizontally homogeneous layers. This allows to solve the vector
radiative transfer equation (VRTE) efficiently. 
For remote sensing applications, the radiance is considered constant
over the instantaneous field-of-view of the instrument and each sensor
element is treated independently in plane-parallel approximation,
neglecting horizontal radiation transport between adjacent pixels
(Independent Pixel Approximation, IPA). In order
to estimate the errors due to the IPA approximation, 
three-dimensional (3D) vector radiative transfer models are required.

So far, only a few such models exist. Therefore, 
the International Polarized Radiative Transfer (IPRT) working group of the
International Radiation Commission (IRC) has initiated a model
intercomparison project in order to provide benchmark results for
polarized radiative transfer. The group has already performed an
intercomparison for one-dimensional (1D) multi-layer test cases
\citep[phase~A, ][]{emde2015}. 
This paper presents the continuation of the intercomparison project
(phase~B) for 2D and 3D test cases: a step cloud, a cubic
cloud, and a more realistic scenario including a 3D cloud field
generated by a Large Eddy Simulation (LES) model and 
typical background aerosols. 

The commonly established benchmark results for 3D polarized
radiative transfer
are available at the IPRT website
(\url{http://www.meteo.physik.uni-muenchen.de/~iprt}). 

\end{abstract}


\end{frontmatter}


\section{Introduction}
\label{sec:intro}

The polarization state of electromagnetic radiation includes
characteristic information about the particles, at which the radiation
has been scattered. This is used by various remote sensing methodologies to
retrieve information about aerosol and cloud optical and microphysical
properties. 

One of the most prominent instruments that measure polarization is the
Polarization and Directionality of the Earth's
Reflectances (POLDER) instrument which was operated onboard the PARASOL (Polarization and
Anisotropy of Reflectances for Atmospheric Sciences coupled with
Observations from a Lidar) satellite \citep{deschamps1994} and has
provided useful observations 
from 2004--2013. The PARASOL mission has demonstrated the usefulness
of multi-spectral directional
polarized measurements. With polarization, it is possible to retrieve
the size distribution of cloud droplets at the top of cloud with high accuracy
\citep{breon2005}.
Further, one can retrieve aerosol optical properties over
bright surfaces such as clouds \citep{waquet2013} or land 
\citep{dubovik2011, xu2017},
which is not possible from unpolarized observations where a dark
background is required.
Moreover, surface properties, i.e. bidirectional polarized reflectance
functions can be derived \citep{maignan2009}. 

The next satellite mission including an instrument similar to POLDER
will be the Multi-Viewing Multi-Channel
Multi-Polarization Imaging mission (3MI) on METOP-SG (Meteorological
Operational Satellite - Second Generation) planned to be launched in
2021 \citep{marbach2015}. Further planned satellite missions with  
polarimeters on board include the Aerosol, Cloud, ocean
Ecosystem (ACE) mission, the MAIA (Multi-Angle Imager for Aerosols)
mission \citep{liu2017}, and the PACE (Plankton, Aerosol, Cloud,
ocean Ecosystem) mission (\url{https://pace.gsfc.nasa.gov/}). 

There are several airborne prototype polarimeters for the preparation
of these missions. The ACE Polarimeter Working Group 
(ACEPWG, \url{https://earthscience.arc.nasa.gov/ACEPWG}) is a
forum for sharing calibration techniques and geophysical parameter
retrieval methods. It also performs intercomparison of the
data collected in field campaigns. Participating instruments are the
airborne prototypes of the Multi-angle SpectroPolarimetric Imager (AirMSPI)
\citep{diner2012}, the Hyper-Angular Rainbow Polarimeter (AirHARP), 
the Spectropolarimeter for Planetary Exploration (AirSPEX)
\citep{van_harten2011}, and
the Research Scanning Polarimeter (RSP) \citep{Cairns1999, Cairns2003}.

AERONET (AErosol RObotic Network) 
is a federation of ground-based remote sensing aerosol networks. It
includes the commercially available ground-based polarimeter, CE318-DP,
developed by CIMEL Electronic (Paris,
France) at many stations. 
 
For correct aerosol retrievals from airborne, satellite or
ground-based polarimetric observations, in particular in partially
cloudy scenes, three-dimensional (3D) radiative transfer models are
required.  \citet{davis2013} studied the influence of 3D effects on 1D
aerosol retrievals from the APS (Aerosol Polarimetry Sensor)
instrument, which was launched with the Glory mission
\citep{mishchenko2007}, but that failed, unfortunately. 
Similarly, 3D effects on aerosol retrievals from 
POLDER observations have been investigated by \citet{stap2016},
\citet{stap2016b} and \citet{cornet2017}.
For aerosol retrievals in partially cloudy scenes,
retrieval errors become significant in the vicinity of clouds. This
effect is also known for aerosol retrievals from unpolarized
observations, e.g. from MODIS \citep{wen2013,varnai2013}.  
An adjoint method for adjusting 3D
atmosphere and surface properties to fit polarimetric measurements has
been derived by \citet{martin2014} and \citet{martin2017}. However, to our knowledge, 
this method has not been applied to real data so far. 
A tomographic cloud
reconstruction method, which uses the 3D SHDOM radiative transfer code
\citep{evans1998} as forward model, has been developed for scalar
multi-angle solar observations \citep{levis2015, levis2017}. The
method has successfully been applied to AirMSPI measurements to retrieve cloud
extinction coefficient, effective droplet size and cloud liquid water
content.   
The influence of 3D effects on polarimetric cloud droplet size
retrievals was investigated by \citet{alexandrov2012} and they found
that 3D effects are negligible, thus the PPA can safely
be used in the forward model of their retrieval algorithm.
\citet{cornet2017} studied the influence of 3D effects on cloud
parameter retrievals from POLDER observations. Their results
confirm that the droplet size retrieval is not much affected by 3D
effects.
\citet{barlakas2016a} and \citet{barlakas2016b} investigated the errors induced
by neglecting horizontal radiation
transport and domain heterogeneities
including polarization in LIDAR-measured dust fields. The differences
in domain-averaged normalized radiances between 
PPA/IPA and 2D calculations are insignificant.
However, in the areas with large spatial variability in optical
thickness, the radiance fields of the
2D calculations differ about 20\% for radiance and polarization 
from the fields of the PPA.
\citet{barlakas2016b} also investigated the error which is induced by neglecting 
polarization in scalar radiative transfer: For pure Rayleigh
scattering errors up to 10.5\% are found for the scalar radiance, in agreement to former
studies (e.g. \citet{Kotchenova2006}).  For 1D and 2D
inhomogeneous atmospheres with Sahara dust aerosols, the maximum error is less
than 1\%, which is explained by relatively high optical thickness und
thus high-order multiple scattering and the asymmetric
scattering phase matrices of the dust particles.

All aforementioned studies used 3D vector radiative transfer
models. So far, all these codes have not been validated regarding 3D
geometry, since established benchmark data does not exist. 
Previous model intercomparisons cover e.g. 3D scalar radiative
transfer (I3RC -- Intercomparison of 3D radiation Codes,
\citep{cahalan2005}), scalar radiative transfer in 1D spherical
geometry \citep{loughman2004}, scalar radiative transfer in the
millimeter/submillimeter spectral region \citep{melsheimer2005}, 
or 1D vector radiative transfer \citep{kokhanovsky2010, emde2015}.
The IPRT intercomparison project presented in this article aims to
provide benchmark results for 3D vector radiative transfer.
The project webpage 
\url{http://www.meteo.physik.uni-muenchen.de/~iprt} provides input
data and results of all models, so that the test cases may be reproduced by
3D vector radiative transfer modelers for validation purposes.

Section~\ref{sec:rt_models} provides an overview of the participating
radiative transfer models. In Section~\ref{sec:definitions}, general
definitions as the coordinate system are
provided. Section~\ref{sec:c1_step_cloud}
presents the model intercomparison for the most simple setup, a step
cloud. In Section~\ref{sec:c2_cubic_cloud}, we present the results for a
cubic cloud. Section~\ref{sec:c3_cumulus} shows the most
realistic simulations for an LES cloud field.  Finally, a brief summary is given in
Section~\ref{sec:summary}.

\section{Radiative transfer models}
\label{sec:rt_models}

\begin{table*}[htbp]
  \centering
  \label{tab:rte_solvers}
  \caption{Overview of radiative transfer models}
  \smallskip
  \begin{tabularx}{1.0\hsize}{lp{3cm}lp{2.2cm}X}
    \hline
    model name & method & geometry & arbitrary \newline output altitude &
    references \\ \hline
    3DMCPOL & Monte Carlo & 1D/3D  & yes &
    \citet{cornet2010,fauchez2014} \\
    MSCART & Monte Carlo & 1D/3D & yes & \citet{wang2017} \\
    MYSTIC & Monte Carlo & 1D/3D$^{(a)}$ & yes & \citet{mayer2009,
                                                 emde2010, emde2016} \\
    SHDOM & spherical harmonics \newline discrete ordinate & 1D/3D  &
                                                                      yes & \citet{evans1998, emde2015} \\
    SPARTA & Monte Carlo & 1D/3D & no & \citet{barlakas2016a, barlakas2016b} \\ \hline
    \multicolumn{5}{l}{$^{(a)}$MYSTIC includes fully spherical
      geometry for 1D and 3D.} \\
  \end{tabularx}
\end{table*}

Table~\ref{tab:rte_solvers} gives an overview of the participating
radiative transfer models. 
In the following, only the MSCART model is briefly described. For short
descriptions of the other models please refer to 
the first publication of the IPRT project \citep{emde2015}.

\subsection{MSCART}
\label{sec:MSCART}

MSCART (Multiple-Scaling-based Cloudy Atmospheric Radiative Transfer)
is a universal simulator of scalar and vector Monte Carlo
radiative transfer in 3D cloudy atmospheres. It is the successor of
the radiative transfer code of MCRT \citep{wang2011b, wang2012}.
MSCART has established a unified forward
and backward scattering order-dependent integral radiative transfer
theoretical framework \citep{wang2017}, which can generalize the
model with variance reduction formalism in a wide range of simulation
scenarios. These include 3D atmospheres with molecules, aerosols
and clouds and 2D surfaces (Lambertian and BPDF (Bidirectional
Polarized reflectance Distribution Function)).

The model is coded in an object-oriented
programming architecture using modern Fortran language
\citep{wang2017}.
This gives the MSCART code a good maintainability and
reusability, and an enhanced capability to add new features in
future. Through several years of development, the model has become a
versatile and sound tool for passive and active remote sensing
applications. In forward mode, it can simulate average radiances over
horizontal areas in specified directions for sunlight and
range-resolved backscattering signals for laser source; in backward
mode, it not only simulates average radiances over horizontal areas in
specified directions but also gives radiances at specified locations
in specified directions for the solar spectral range.
Polarization has been 
recently added to the model for macroscopically isotropic and
mirror-symmetric scattering medium.

More importantly, 
sophisticated variance reduction techniques are implemented into the
model to speedup simulations for cloudy atmospheres with highly
forward-peaked scattering. The previous studies reduce variance either
by using the scattering phase matrix forward truncation technique or
the target directional importance sampling technique. In this model, a
novel scattering order-dependent variance reduction method is used to
combine both of them and a new scattering order sampling algorithm is
implemented to achieve an order-dependent tuning parameter
optimization \citep{wang2017}. The MSCART software package with
several simulation examples can be freely downloaded after
registration from the designated website
\url{http://mscart.nuist.edu.cn}.

\section{General definitions}
\label{sec:definitions}

\subsection{Model coordinate system and Stokes vector}

For all test cases, the four 
Stokes parameters \citep{chandrasekhar50, hansen1974, 
  mishchenko2002, wendisch2012} are calculated:
\begin{eqnarray}
  {\bf I} =
  \begin{pmatrix}
    I \\ Q \\ U \\ V
  \end{pmatrix} 
  = \frac{1}{2}\sqrt{\frac{\epsilon}{\mu_p}}
  \begin{pmatrix}
    E_\parallel E_\parallel^\ast + E_\perp E_\perp^\ast \\
    E_\parallel E_\parallel^\ast - E_\perp E_\perp^\ast \\
    -E_\parallel E_\perp^\ast - E_\perp E_\parallel^\ast \\
    i(E_\parallel E_\perp^\ast- E_\perp E_\parallel^\ast) 
  \end{pmatrix}
  \label{eq:stokes}
\end{eqnarray}
Here, $E_\parallel$ and $E_\perp$ are the components of the electric field vector
parallel and perpendicular to
the reference plane, respectively. The pre-factor on the right hand
side contains the electric permittivity $\epsilon$ and the magnetic
permeability $\mu_p$.

The degree of polarization $P$ is calculated from the Stokes vector as follows: 
\begin{equation}
  \label{eq:pol_deg}
  P=\frac{\sqrt{(Q^2+U^2+V^2)}}{I}
\end{equation}

\begin{figure}
  \centering
  \includegraphics[width=.9\hsize]{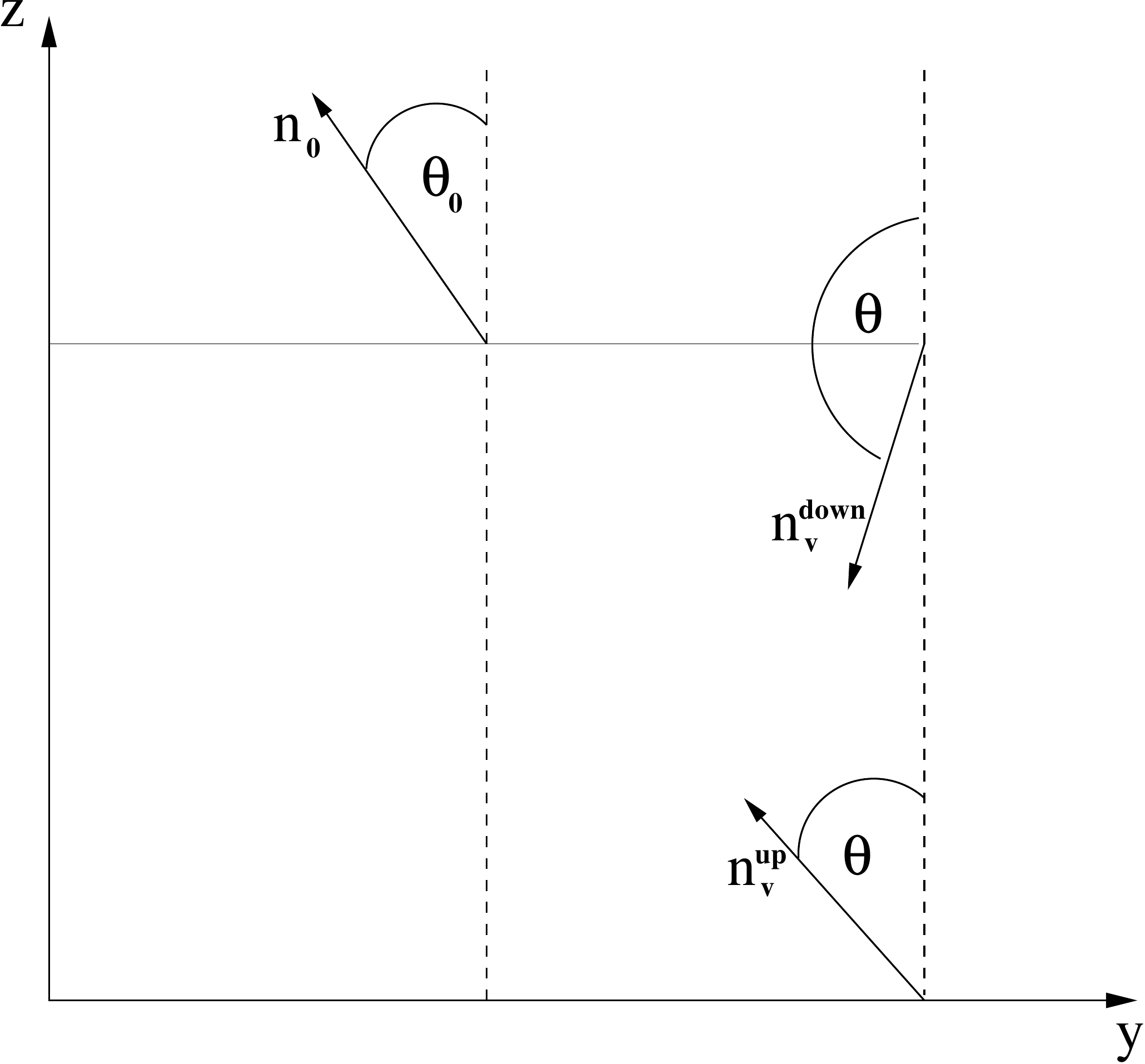}
  \caption{Definition of viewing zenith angle $\theta$ and solar
    zenith angle $\theta_0$ for up- and down-looking directions.}
  \label{fig:coordinates}
\end{figure}
The model coordinate system is defined by the vertical (z-axis), the Southern
direction (x-axis) and the Eastern direction (y-axis).
The Stokes vector is defined in the reference frame spanned by the
z-axis and the propagation direction of the radiation. 
The sign of Stokes parameters $U$ and $V$ depends on the definition of
the model coordinate system. The results shown in this paper are for
the coordinate system as defined in the books by
\citet{hovenier2004} and \citet{mishchenko2002}. The sign of $U$ and
$V$ changes when the viewing azimuthal angle definition is changed from
anti-clockwise to clockwise and also when the definition of the
viewing zenith angle is with respect to the downward normal instead of
the upward normal. 
The models SHDOM and 3DMCPOL use the definition according to
\citet{hovenier2004}. SPARTA uses a
different coordinate system but the signs are consistent with
\citet{hovenier2004}. MYSTIC and MSCART also use different coordinate
systems and obtain opposite signs for $U$ and $V$, all results for
these Stokes components shown
in this paper have been multiplied by -1.
  
The position of the sun is defined by the vector pointing from the
surface to the sun position ${\bf n}_0$ (see
Figure~\ref{fig:coordinates}). The solar zenith angle $\theta_0$ is
therefore defined in the range from 0\degree\ to
90\degree. Twilight conditions, for which the solar zenith angle
is larger than 90\degree, are not considered in this
intercomparison because we neglect the sphericity of the Earth
in this study. 
The viewing zenith angle $\theta$ is between 0\degree\ to 90\degree\ for observer
positions at the surface looking upwards into direction ${\bf
  n}_v^{up}$, and between 90\degree\ and
180\degree\ for observer positions at the top of the
atmosphere looking downwards into direction ${\bf n}_v^{down}$.
When the solar azimuth angle $\phi_0$
equals the viewing azimuth angle $\phi$ for the observer at the surface, the
viewing direction is towards the sun. For an observer at the top of
the atmosphere, the sun is in the back of the observer when
$\phi-\phi_0$=180\degree.

\subsection{Statistics}
\label{sec:stats}

In order to compare the models quantitatively, we calculate the
mean radiance for each test case:
\begin{equation}
  \label{eq:mean_def}
  I_{\rm mean,j} = N_{\rm pixels}^{-1} \sum_{i=1}^{N_{\rm pixels}} \left|I_{i,j}\right|
\end{equation}
Here, $j$ denotes the index of the Stokes vector component. 
Note that since $Q$, $U$, and $V$ can be positive or
negative, we calculate the mean of the absolute values for comparison
of the model results.
We also calculate the
mean relative standard deviation 
\begin{equation}
  \label{eq:rel_std_def}
  \sigma_{\rm rel,j} = \frac{\sum_{i=1}^{N_{\rm pixels}}
    \sigma(I_{i,j})}{\sum_{i=1}^{N_{\rm pixels}}\left| I_{i,j} \right|}
\end{equation}
where $\sigma(I_{i,j})$ is the standard deviation of the Stokes vector
component $j$ at pixel $i$.   
Further we compute the relative root mean square differences
\begin{equation}
  \label{eq:rms_def}
  \Delta_{\rm RMS,j} = \frac{\sqrt{\sum_{i=1}^{N_{\rm pixels}}
      \left(I_{i,j}-I_{i,j}^{\rm ref}\right)^2}}
  {\sqrt{\sum_{i=1}^{N_{\rm pixels}}{I_{i,j}^{\rm ref}}^2}}
\end{equation}
Here, $I_{i,j}^{\rm ref}$ are the results of the reference model.

For test cases C2 and C3, we also calculate a match fraction $q_j$ for
each Stokes component $j$ which we
define as follows: 
we calculate the number of pixels $N_{\rm match,j}$ for which the absolute
difference between a model and the reference model is smaller than the
sum of two standard deviations (2$\sigma$) of both models 
\begin{equation}
  \label{eq:Nmatch}
  \left|I_{i,j} - I_{i,j}^{\rm ref} \right| <
  2 \left(\sigma(I_{i,j}) +  \sigma(I_{i,j}^{\rm ref}) \right) 
\end{equation}
We take into account all pixels $N_{\rm all,j}$ with radiance values greater than
10$^{-8}$. 
The match fraction is
\begin{equation}
  \label{eq:match_fraction}
  q_j=\frac{N_{\rm match,j}}{N_{\rm all,j}}
\end{equation}
When the two models agree within their statistical noise quantified by
two standard deviations of each pixel, the match
fraction should be larger than 95.45\% according to Gaussian
statistics.  
The deterministic code SHDOM is not noisy and therefore does not calculate
a standard deviation, thus we only calculate $\Delta_{\rm RMS,j}$.

\section{Test case C1 -- Step cloud}
\label{sec:c1_step_cloud}
 
\subsection{C1 -- Model setup}
\label{sec:c1_step_cloud_setup}
\begin{figure}
  \centering
  \includegraphics[width=1.\hsize]{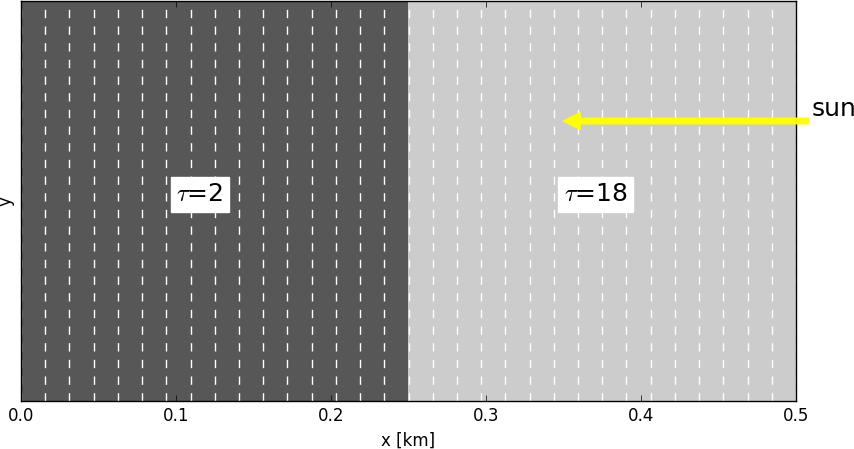}
  \caption{Definition of step cloud, test case C1. In x-direction the optical thickness of the cloud changes periodically between 2 and 18. In y-direction the cloud extends infinitely.}
  \label{fig:c1_step_cloud_def}
\end{figure}

The first test scenario corresponds to a simple step cloud with 
32 pixels along the x-direction (see Figure~\ref{fig:c1_step_cloud_def}). 
The first 16 pixels have an optical
thickness $\tau$ of 2 and the remaining 16 pixels $\tau=$18.
The size of the cloud field is 0.5~km and  
all pixels have a width of 0.5/32~km\,=15.625~m.
The geometrical thickness of the cloud along the z-axis is 0.25~km and 
the cloud extends in y-direction infinitely. We assume periodical
boundary conditions in x-direction (for Monte Carlo models this means
that photons that leave the domain
on one side of the domain at a certain vertical position 
re-enter the domain on the opposite side at the same vertical position
without changing their propagation direction). Periodical boundary
conditions are often used in Monte Carlo radiative transfer codes,
because they ensure energy conservation and are appropriate for most
applications, given that the model domain is sufficiently large so
that the region of interest is not affected by the edge effects
\citep{mayer2009}. 
The provided cloud optical properties were calculated using Mie
theory for a wavelength
of 800~nm using a gamma size distribution:
\begin{equation}
  n(r)=C r^\alpha \exp\left(-\frac{(\alpha+3)r}{r_{\rm eff}}\right)
\end{equation}
Here $\alpha$ was set 7 and the effective radius $r_{\rm eff}$ was
10~$\mu$m. The effective variance is $v_{\rm
  eff}=\frac{1}{\alpha+3}=0.1$ which is a typical value for a liquid water cloud.
The constant $C$ is obtained by normalization.
The refractive index of water at 800~nm is 1.325+1.250\ee{-7}\,{\sl
  i} \citep{segelstein1981}, this means that there is
almost no absorption. 
The surface albedo was set to 0 for this test case, i.e. we have a black
surface corresponding to a water surface outside the sunglint region.  
The solar azimuth angle $\phi_0$ is 0\degree\ (see also
Figure~\ref{fig:c1_step_cloud_def}). 

Ten different sun-observer geometries as shown 
in Table~\ref{tab:c1_settings} were calculated. Each geometry is
defined by  the solar zenith angle $\theta_0$, the solar azimuth angle
$\phi_0$=0\degree, the altitude of the observer, the viewing zenith
angle $\theta$ and the viewing azimuth angle $\phi$.  
The table also includes the scattering
angle $\theta_s=\arccos({\bf n}_0\cdot{\bf n}_v)$, where ${\bf n}_0$
is the sun position vector and ${\bf n}_v$ is the viewing
direction vector.
The angle between sun direction and viewing direction is
the scattering angle $\theta_s$ for single scattering. 
\begin{table}
  \centering
  \begin{tabular}{l l l l l l}
    \hline 
    no. & $\theta_0$ [\degree] & $z$ [km] & $\theta$ [\degree] & $\phi$ [\degree] & $\theta_s$ [\degree]\\ \hline
    1 & 0 & 0 & 60 & 0 & 60 \\
    2 & 60 & 0 & 0 & 0 & 60 \\  
    3 & 60 & 0 & 30 & 0 & 30\\
    4 & 60 & 0 & 30 & 180 & 90 \\        
    5 & 0 & 0.25 & 180 & 0 & 180 \\ 
    6 & 0 & 0.25 & 140 & 0 & 140 \\
    7 & 0 & 0.25 & 120 & 0 & 120 \\ 
    8 & 60 & 0.25 & 180 & 0 & 120 \\
    9 & 60 & 0.25 & 120 & 0 & 60 \\
    10 & 20 & 0.25 & 120 & 135 & 132.8\\ \hline
  \end{tabular}
  \caption{Geometrical settings for scenario C1 -- step cloud.  The
    solar azimuth angle is $\phi_0$=180\degree\ for all cases. 
}
  \label{tab:c1_settings}
\end{table} 
For each of the directions we calculate the pixel averaged Stokes
vector for all 32 pixels in the model domain.
All Monte Carlo simulations were supposed to be run with 10$^7$
``photons'' \footnote{We use the term
  ``photon'' to represent an imaginary discrete amount of
  electromagnetic energy transported in a specific direction. It is
  not related to the QED photon \citep{mishchenko2014}.} per
pixel, i.e. 32\ee{7}\ photons in total.  
Two model runs with more photons were performed to obtain a reference.
Table~\ref{tab:c1_models} includes more details about the settings of the Monte Carlo
models. The MSCART model was run in forward tracing mode (MSCART-F)
and backward tracing mode (MSCART-B).
For SHDOM the number of discrete
ordinates in zenith and azimuth angle were $N_{\mu}=64$ and $N_{\phi}$=128,
and the spatial resolution was $N_{x}$=160 and $N_{z}$=81 for the base grid
with cell splitting accuracy of 0.002.

\begin{table}
  \centering
  \begin{tabular}{l l l l l}
    model & $N_{ph}$ & VR & TM \\ \hline 
    3DMCPOL & 32\ee{7} & no & F \\
    SPARTA & 32\ee{7} & no & F \\
    SPARTA 10$^{11}$ & 10$^{11}$ & no & F \\
    MSCART-F & 32\ee{7} & yes & F \\
    MSCART-B & 32\ee{7} & yes & B \\
    MYSTIC & 32\ee{7} & yes & F \\
    MYSTIC-NV & 32\ee{7} & no & F \\
    MYSTIC 32\ee{8} & 32\ee{8} & yes & F \\
  \hline 
  \end{tabular}
  \caption{Monte Carlo model settings for scenario C1 -- step
    cloud. $N_{ph}$ is the number of photons, the VR column shows whether
    the models used variance reduction methods for spiky scattering phase
    functions and TM gives the tracing method (forward tracing (F) from
    the sun towards observer or backward (B) tracing from observer towards
    the sun).}
  \label{tab:c1_models}
\end{table}

\subsection{C1 -- Results}
\label{sec:c1_step_cloud_results}

\begin{figure*}[p]
  \centering
  \includegraphics[width=.8\hsize]{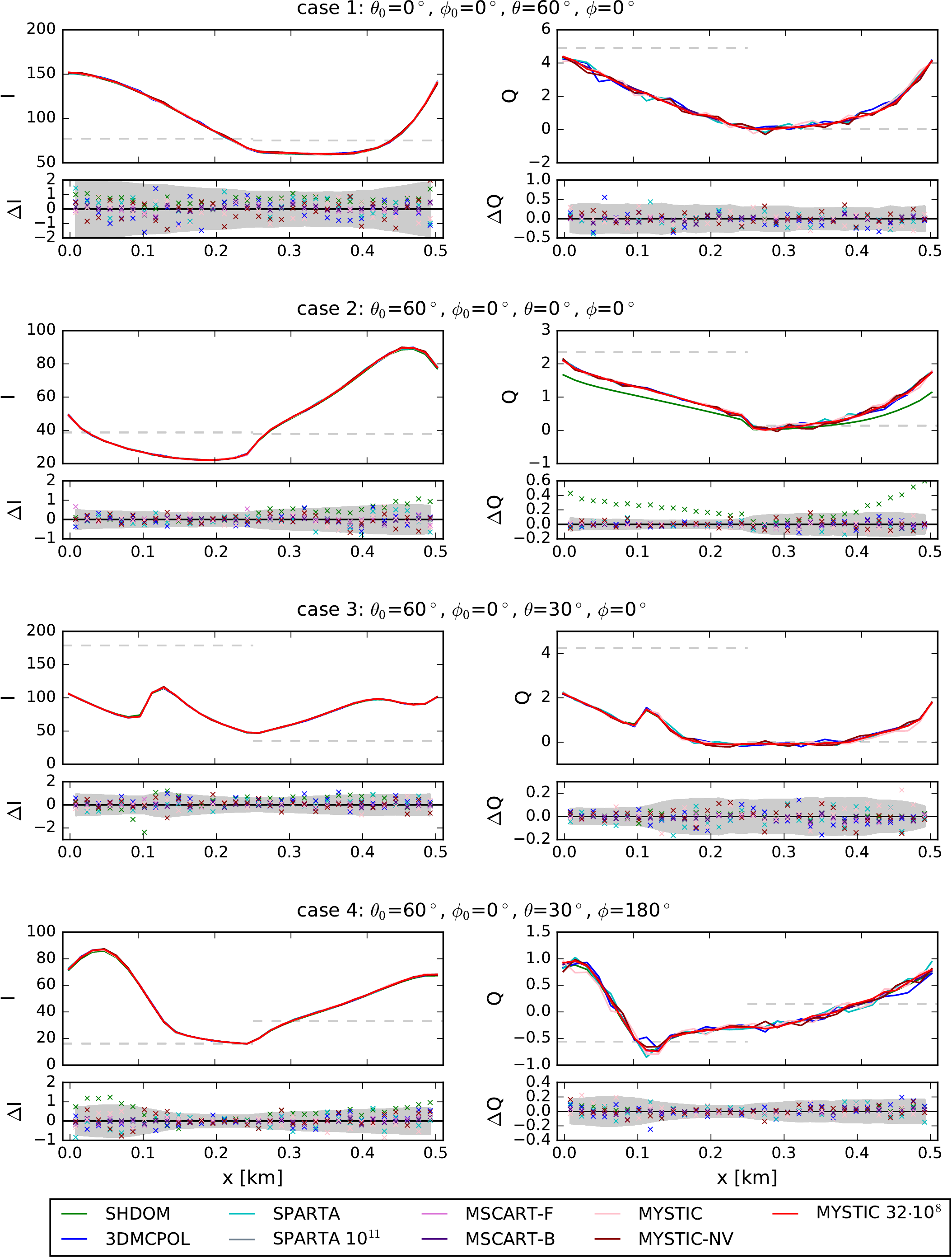}
  \caption{Stokes vector components $I$ (left) and $Q$ (right) for the
    step cloud scenario C1. Here, the observer is placed at the bottom
    of the cloud layer at position $x$ and it looks upwards, its
    viewing direction is given by ($\theta$,$\phi$). The solar zenith
    angle is 0$^\circ$ for case 1 and 60$^\circ$ for cases 2--4.  The
    solar azimuth angle is 0$^\circ$ for all cases. The small panels
    show the absolute differences between various models (see legend)
    and accurate MYSTIC simulations obtained with 10$^8$ photons/pixel
    (MYSTIC 32$\cdot$10$^8$). The grey range corresponds to 2$\sigma$ of MYSTIC
    simulations without variance reduction methods and with 10$^7$
    photons/pixel (MYSTIC-NV). The Stokes vector components are
    normalized to 1000/E$_0$. All cases are in the solar principal
    plane where the Stokes components $U$ and $V$ are exactly 0 and,
    therefore, not shown. The grey dashed lines show 1D independent pixel
    calculations. }
  \label{fig:c1_results_up}
\end{figure*}
Figure~\ref{fig:c1_results_up} shows the results for up-looking geometries (test cases
1--4). The observer is placed at the bottom of the cloud layer at
position $x$ looking upwards. In the plots, the Stokes vector
components are normalized to 1000/$E_0$, where $E_0$ corresponds to
the extraterrestrial irradiance at the simulated wavelength (here
800\,nm). The models generally produce the same spatial radiance
patterns, meaning that they all handle horizontal photon transport correctly.
The Monte Carlo models without variance reduction show
significant noise in particular for $Q$ which is relatively small below
the cloud. 

Note that
using a plane-parallel (1D) model and IPA, we would get only two different values
in each of the plots, one from $x$=0\,km to $x$=0.25\,km where the optical
thickness of the cloud is 2, and one from $x$=0.25\,km to $x$=0.5\,km where
the optical thickness is 18. IPA results are
included as grey dashed lines in the plots. Comparing them with the 3D
results shows that in scenarios similar to the step cloud case,
i.e. highly variable optical thickness on a small scale
within the domain, it is important to consider horizontal photon
transport, for the total radiance and also for the polarized
radiance.

The small panels show the absolute
differences between the model simulations (see legend) and accurate MYSTIC
simulations obtained with 10$^8$ photons/pixel (MYSTIC 32$\cdot$10$^8$). The grey
range corresponds to 2 standard deviations (2$\sigma$)
of MYSTIC simulations without variance
reduction methods and with 10$^7$ photons/pixel (MYSTIC-NV), which
corresponds to the default Monte Carlo model settings for this scenario.
The differences for the Monte Carlo codes are within the expected
2$\sigma$ range. Some systematic deviations are found for the only
deterministic code SHDOM: for case 2, where the sensor looks exactly into
zenith direction, we see rather large deviations for
Stokes component $Q$. The IPRT 1D intercomparison \citep{emde2015} 
showed a problem SHDOM has for Stokes components $Q$ and $U$ near
$\theta$=0\degree\ and $\theta$=180\degree\ 
for highly peaked phase functions, and this issue has
not yet been resolved.
 
\begin{figure*}[p]
  \centering
  \includegraphics[width=.75\hsize]{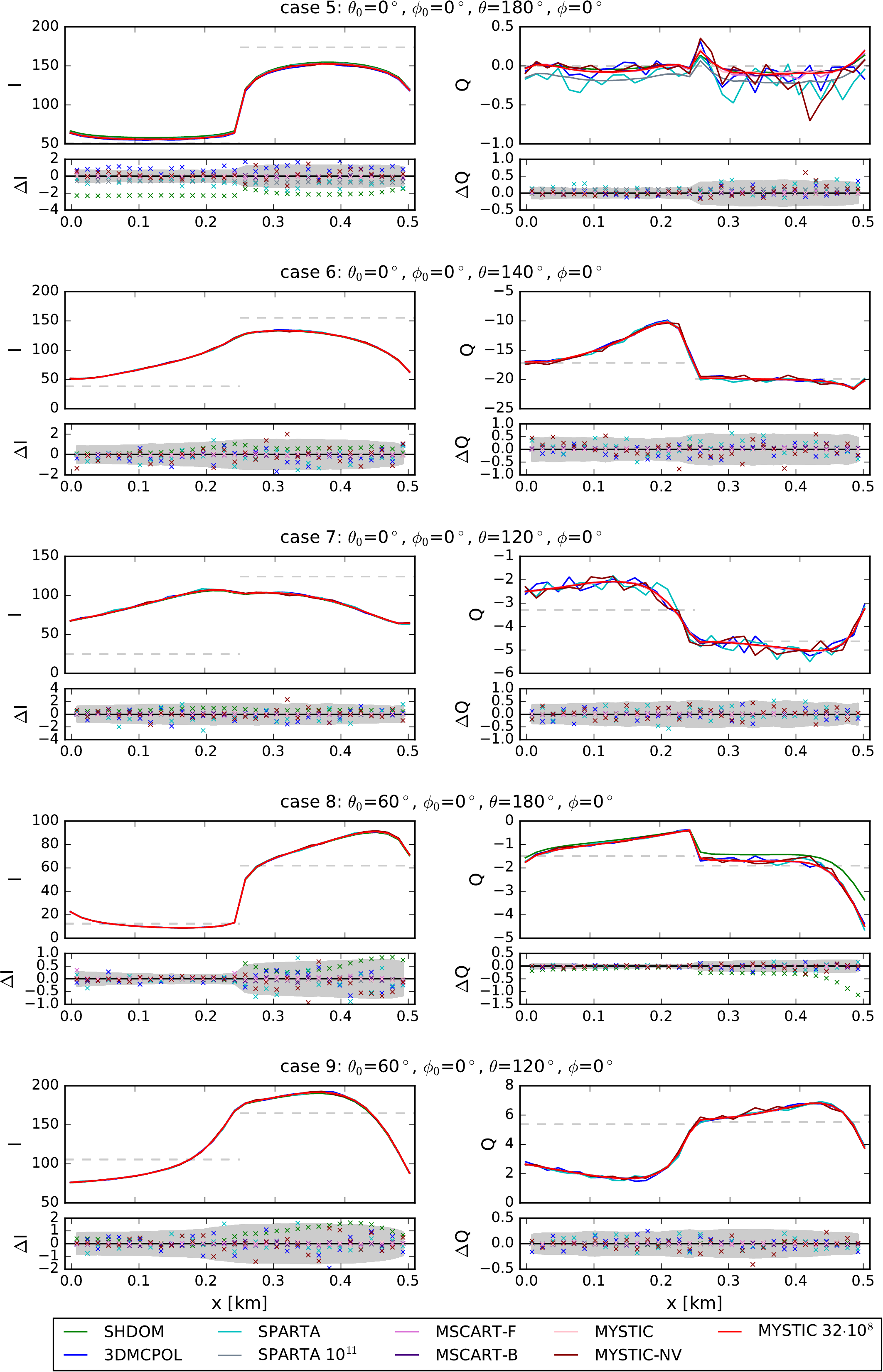}
  \caption{Stokes vector components I (left) and Q (right) for the
    step cloud scenario C1. Here the observer is placed at the top of
    the cloud layer at position x and it looks downwards and its viewing
    direction is given by ($\theta$,$\phi$). The solar zenith angle is
    0$^\circ$ for cases 5--7 and 60$^\circ$ for cases 8--9. The solar
    azimuth angle is 0$^\circ$ for all cases. The small panels show
    the absolute differences between various models (see legend) and
    accurate MYSTIC simulations obtained with 10$^8$ photons/pixel (MYSTIC
    32$\cdot$10$^8$). The grey range corresponds to 2$\sigma$ of MYSTIC
    simulations without variance reduction methods and with 10$^7$
    photons (MYSTIC-NV). The Stokes vector components are normalized
    to 1000/E$_0$. All cases are in the solar principal plane where
    the Stokes components $U$ and $V$ are exactly 0 and, therefore, not
    shown. The grey dashed lines show 1D independent pixel
    calculations.}
   \label{fig:c1_results_down}
\end{figure*}
Figure~\ref{fig:c1_results_down} shows the results for down-looking geometries (test cases
5--9), where the observer is placed at the top of the cloud layer at
position $x$ looking downwards. Again, we find generally a good agreement
between all models, i.e. the differences are within the 2$\sigma$
range of MYSTIC-NV. There are only two obvious problems: (1) $Q$ for case 5,
where the sensor looks exactly nadir and the sun is in zenith,
is biased towards a negative
value for SPARTA. (2) $Q$ for case 8, where the sensor
also looks exactly nadir but the solar zenith angle is 60\degree, is
positively biased for SHDOM.

As for the up-looking cases, models without variance reduction
show significant noise, especially for case 5 and case 7, where $Q$ is
very small. 

\begin{figure*}
  \centering
  \includegraphics[width=1.0\hsize]{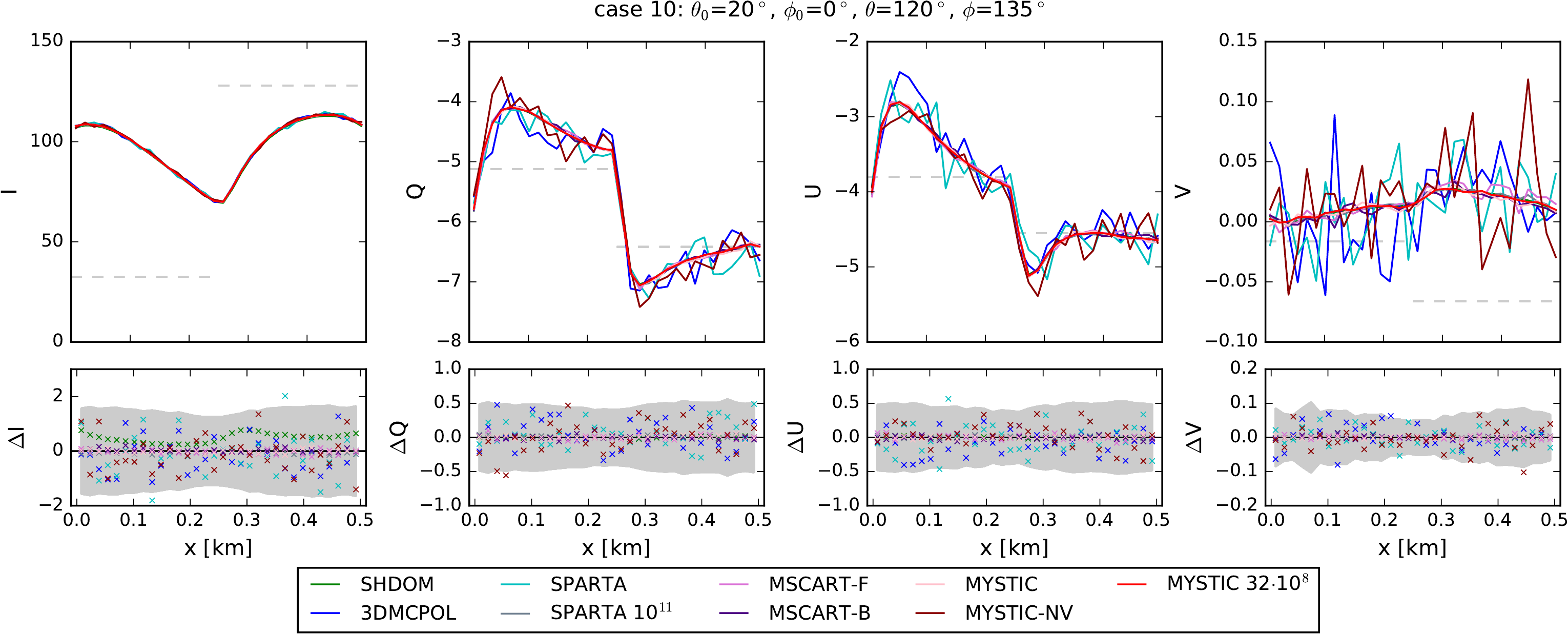}
  \caption{Stokes vector components I, Q, U, and V for the step cloud
    scenario C1. Here, the observer is placed at the top of the cloud
    layer at position x and it looks downwards and its viewing direction
    is given by ($\theta$=120$^\circ$,$\phi$=135$^\circ$). The solar
    zenith angle is $\theta_0$=20$^\circ$ and the solar azimuth angle
    is $\phi_0$=0$^\circ$. This sun-observer geometry is outside the
    solar principal plane and, hence, non-zero values for $U$ and
    $V$ are expected. The small panels show the absolute differences between
    various models (see legend) and accurate MYSTIC simulations
    obtained with 10$^8$ photons/pixel (MYSTIC 32$\cdot$10$^8$). The grey range
    corresponds to 2$\sigma$ of MYSTIC simulations without variance
    reduction methods and with 10$^7$ photons (MYSTIC-NV). The Stokes
    vector components are normalized to 1000/E$_0$. The grey dashed lines
    show 1D independent pixel calculations.}
   \label{fig:c1_results_case10}
\end{figure*}
Figure~\ref{fig:c1_results_case10} shows the results for case 10, where 
the observer is placed at the top of the cloud
layer at position $x$ looking downwards with viewing direction
given by ($\theta$=120$^\circ$,$\phi$=135$^\circ$). The solar
zenith angle is $\theta_0$=20$^\circ$ and the solar azimuth angle
is $\phi_0$=180$^\circ$. This sun-observer geometry is the only one 
outside the
solar principal plane, hence, we expect non-zero values for $U$ and
$V$. The models generally agree within the 2$\sigma$ range of
MYSTIC-NV. However, for $V$ the noise is very large and only models with
variance reduction (MYSTIC and MSCART), or with a very large number of
photon runs (SPARTA 10$^{11}$) produce a significant curve, where the
noise is not larger than the values of $V$. The achieved accuracy of $V$ is
consistent with \citet{garciamunoz2015b}, who shows that 10$^9$ photons
or more are required to resolve circular polarization with values of
$V/I$ of the order of 10$^{-5}$.
A small bias in $I$ ($\Delta I/I \approx$0.5\%) is seen for SHDOM. 

In order to compare the models quantitatively, we calculate the
statistical quantities as defined in Section~\ref{sec:stats}. 
As reference model we have taken MYSTIC 32\ee{8}. 
\begin{figure*}
  \centering
  \includegraphics[width=0.9\hsize]{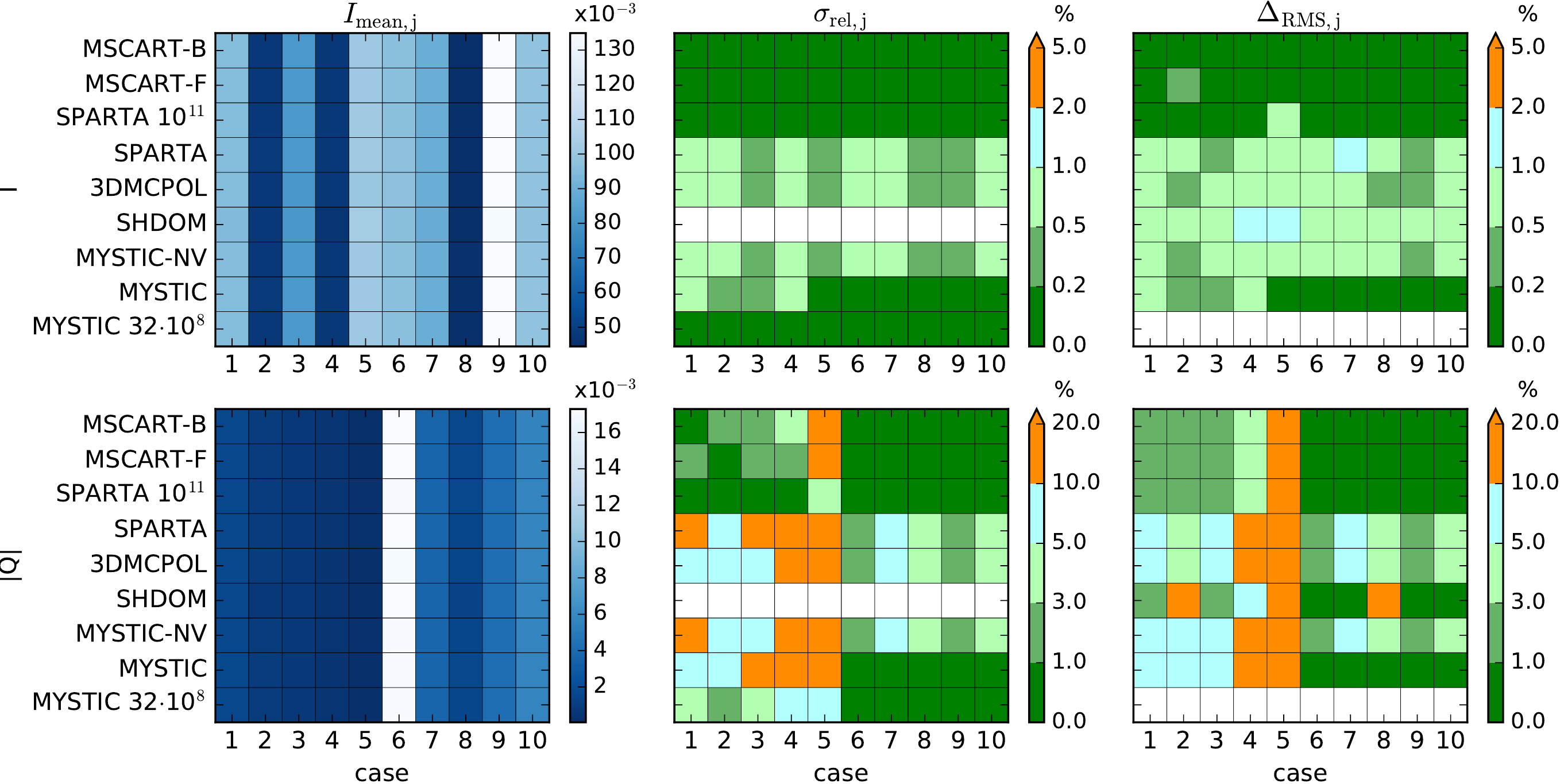}
  \caption{Statistics of Stokes vector components I and Q for the step
    cloud scenario C1. The left panels show the mean radiance (for Q
    the mean of the absolute values) for all
    models and all 10 cases. The middle panels show the
    standard deviations and the right panels show the root mean
    square differences $\Delta_{\rm RMS}$ in per cent. 
    U and V are not shown here because they are 0
    except for case 10.} 
  \label{fig:c1_stats}
\end{figure*}
Results are shown in Figure~\ref{fig:c1_stats}. The left panels show the
mean radiances: rows correspond to the test cases 1--10 and
columns correspond to results of the different models. We see a pattern of vertical
lines which means that the mean radiances are approximately the same
in all models. The middle plots show the mean relative standard
deviations. For the default settings with 32\ee{7} photons without
using variance reduction methods (SPARTA, 3DMCPOL, MYSTIC-NV) 
$\sigma_{\rm rel,I}$ is between 0.2\% and 1\% (for the first Stokes
component $I$). For up-looking cases 1--4, $\sigma_{\rm rel,I}$ is
approximately the same for MYSTIC with and without variance
reduction (MYSTIC-NV). This shows that for up-looking directions the variance
reduction method VROOM \citep{buras2011} 
as implemented in MYSTIC does not reduce the noise as
expected and it
should be optimized. For MSCART, $\sigma_{\rm rel,I}$ is below 0.2\%
for all cases in forward and backward tracing mode. It is also
below 0.2\% for the SPARTA run with 10$^{11}$ photons and the MYSTIC run with
32\ee{8} photons, which is the
expected result since $\sigma\propto N_{\rm ph}^{-1/2}$.   
For $Q$, the relative standard deviation is significantly larger
because the mean values of $Q$ are 1-2 orders of magnitude smaller
than $I$. For Monte Carlo runs with default settings, $\sigma_{\rm
  rel,Q}$ is between 1\% and 10\% for down-looking cases 6--10. For
up-looking cases and case 5, the magnitude of $Q$ is very small and  
$\sigma_{\rm rel,Q}$ ranges from 5\% to more than 20\%. With variance
reduction, $\sigma_{\rm rel,Q}$ is reduced to values below 1\% for
cases 5--10. The MSCART variance reduction method reduces $\sigma_{\rm
  rel,Q}$ to values below 3\% also for up-looking directions. For $Q$,
we also find that the MYSTIC variance reduction does not reduce the
noise as
expected for up-looking directions. 
SHDOM does not calculate standard deviations, thus the row
corresponding to SHDOM results is left white.

The right panels show the relative RMS differences $\Delta_{\rm
  RMS,j}$. These should be similar to the $\sigma_{\rm rel,j}$ if the
model results are not biased compared to the reference model. 
Indeed, we obtain similar values. For
up-looking cases 1--4, $\Delta_{\rm RMS,Q}$ is larger than
$\sigma_{\rm rel,Q}$ for the accurate models MSCART-F, MSCART-B and
SPARTA 10$^{11}$, the reason for this is that the RMS difference is
dominated by the less accurate reference calculation MYSTIC
32\ee{8}. For SHDOM, $\Delta_{\rm RMS,I}<$1\% for all cases except 4
and 5, for those  $\Delta_{\rm RMS,I}<$2\%. $\Delta_{\rm
  RMS,Q}$ of SHDOM is smaller than the relative standard deviation of MYSTIC
32\ee{8} for all cases except 2, 5, and 8, which
are for either nadir or zenith viewing geometry.

\section{Test case C2 -- Cubic cloud}
\label{sec:c2_cubic_cloud}

\subsection{C2 -- Model setup}
\label{sec:c2_cubic_cloud_setup}

\begin{figure}
  \centering
  \includegraphics[width=0.9\hsize]{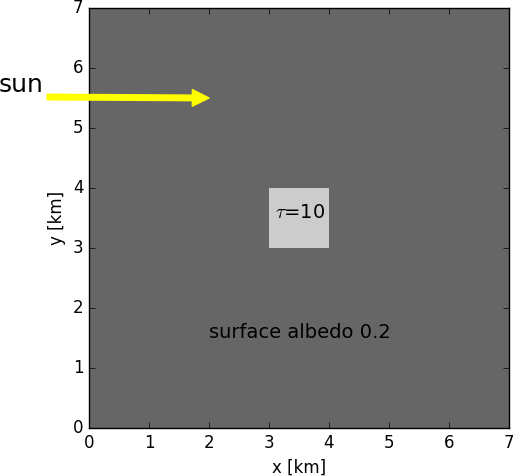}
  \caption{Definition of cubic cloud, test case C2. }
  \label{fig:c2_cubic_cloud_def}
\end{figure}
The second test case includes a 1$\times$1$\times$1\,km$^3$ cubic cloud (see Figure~\ref{fig:c2_cubic_cloud_def}).
The domain size is 7$\times$7\,km$^2$ in the x-y-plane and 5\,km in
z-direction. The cloud is located in the domain center: In x and y
directions it extends from 3--4\,km, respectively,
and in z-direction it extends
from 2--3\,km. We assume periodic boundary conditions in x and
y-directions. The vertical optical thickness of the cloud is 10 and,
within the cube, the cloud extinction coefficient is constant. We use the
same cloud optical properties as for the step cloud: a
wavelength of 800\,nm and and an effective droplet radius of 10\,$\mu$m. Here, we
assume that the cloud droplets do not absorb, i.e., the single
scattering albedo was set to 1.0. We include a Lambertian surface with
albedo 0.2 at the bottom of the domain. 
The solar azimuth angle is 180\degree, which
means that the sun direction is along the 
x-axis (photons enter the atmosphere in positive x-direction).
We calculate the polarized radiation fields with a
spatial resolution of 70$\times$70~pixels for various viewing
directions (see Table~\ref{tab:c2_settings}). Surface radiances are
calculated for $\theta_0$\,=\,20\degree\ and top-of-domain radiances
for $\theta_0$\,=40\degree. Table~\ref{tab:c2_settings} also includes
the scattering angle $\theta_s=\arccos({\bf n}_0\cdot{\bf n}_v)$ for
all of the 9 test cases.
\begin{table}
  \centering
  \begin{tabular}{l l l l l l}
    \hline 
    no. & $\theta_0$ [\degree] & $z$ [km] & $\theta$ [\degree] & $\phi$
                                                               [\degree]
    &  $\theta_s$ [\degree]\\ \hline
    1 & 20 & 0 & 40 & 0 & 60.0 \\
    2 & 20 & 0 & 40 & 60 & 52.4\\  
    3 & 20 & 0 & 40 & 120 & 33.9 \\
    4 & 20 & 0 & 40 & 180 & 20.0 \\        
    5 & 40 & 5 & 180 & 0 & 140.0\\ 
    6 & 40 & 5 & 140 & 0 & 180.0 \\
    7 & 40 & 5 & 140 & 60 & 142.5 \\ 
    8 & 40 & 5 & 140 & 120 & 112.3 \\
    9 & 40 & 5 & 140 & 180 & 100.0 \\ \hline
      \end{tabular}
  \caption{Geometrical settings for cases C2 (cubic cloud) and C3
    (cumulus cloud). The solar azimuth angle is $\phi_0$=180\degree\
    for all cases.}
  \label{tab:c2_settings}
\end{table}

Two sets of simulations are performed: first, the cubic cloud is in vacuum
and second, the cloud is included in a homogeneous Rayleigh scattering
layer from 0 to 5~km  with optical thickness set to 0.5 (this is a
typical value at a wavelength of about 370~nm for the Earth atmosphere).
The Rayleigh depolarization
factor is set to 0. For the exact definition of the Rayleigh phase
matrix, the reader is referred to \citet{emde2015}.

\begin{figure*}
  \centering
  \includegraphics[width=0.7\hsize]{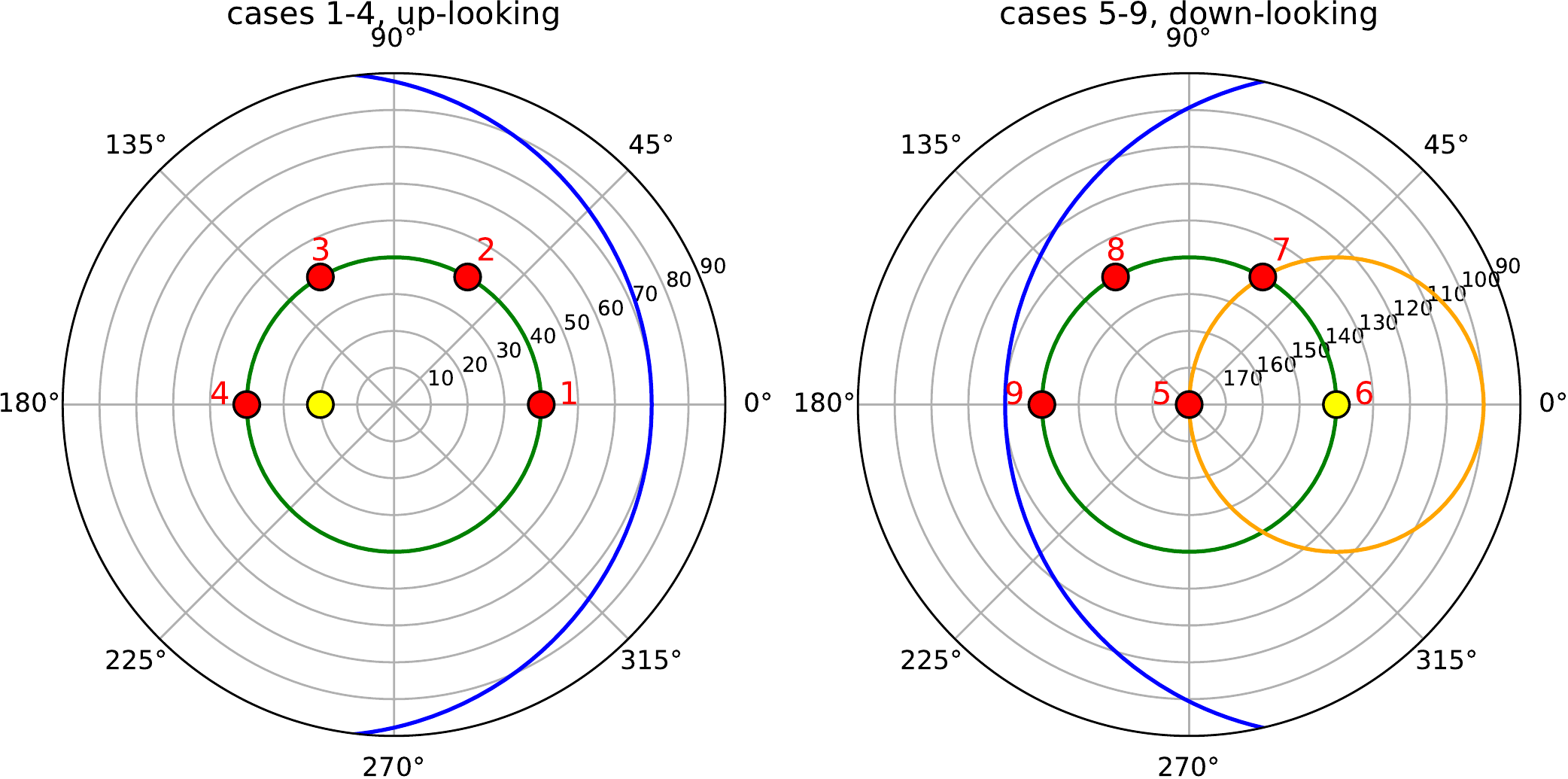}
  \caption{Viewing geometry for scenarios C2 (cubic cloud) and C3
    (cumulus cloud). (Left panel:) up-looking directions
    (cases 1--4). The sun position (yellow circle) is $(\theta_0,
    \phi_0)=(20^\circ,180^\circ)$. The viewing zenith angle is
    constant (40\degree, green circle). The blue line corresponds to
    viewing directions
    with a scattering angle of 90\degree. (Right panel:) down-looking
    directions, cases 5--9. The sun in in the back  
    from the perspective of the down-looking
    observer at the top of the atmosphere, i.e. the position vector of
    the sun is opposite to the viewing direction vector. The yellow
    circle at $(140^\circ,0^\circ)$ corresponds to the sun position  
    $(\theta_0, \phi_0)=(40^\circ,180^\circ)$. The viewing
    angle is nadir for case 5 and 140\degree\ otherwise (cases 6--9,
    green circle). The blue line shows viewing directions with a scattering
    angle of 90\degree\ and the orange circle viewing directions with a
    scattering angle of 140\degree. } 
  \label{fig:c2_geometry}
\end{figure*}
The viewing geometries are illustrated in
Figure~\ref{fig:c2_geometry}. The left plot shows the up-looking
directions 1--4 and the right plot the down-looking directions
5--9. For up-looking directions in clear-sky regions, the highest
degree of polarization is expected at a scattering angle of 90\degree\
(blue line) because for this angle Rayleigh scattering causes the
maximum degree of polarization of 100\% for single
scattering. Therefore, in the clear-sky region we expect decreasing
polarization from case 1 to case 4. Also for down-looking directions
we expect in clear-sky regions the highest degree of polarization at a
scattering angle of 90\degree. Since case 9 is closest to the blue line,
we expect for this case the highest polarization in the clear sky
region. At a scattering angle of 140\degree\ (orange line, cloud- or rainbow
region), clouds cause high polarization, and therefore, in the cloudy region
we expect the highest polarization for cases 5 and 7.

\begin{table}
  \centering
  \begin{tabular}{l l l l l}
    model & $N_{ph}$ & VR & TM \\ \hline 
    3DMCPOL & 49\ee{9} & no & F \\
    SPARTA & 1\ee{11} & no & F \\
    MSCART-F & 49\ee{9} & yes & F \\
    MSCART-B & 49\ee{9} & yes & B \\
    MYSTIC & 49\ee{9} & yes & F \\
    MYSTIC-B & 49\ee{9} & yes & B \\
  \hline 
  \end{tabular}
  \caption{Monte Carlo model settings for scenario C2 -- cubic
    cloud. $N_{ph}$ is the number of photons, the VR column shows whether
    the models used variance reduction methods for spiky scattering phase
    functions and TM gives the tracing method (forward tracing (F) from
    the sun towards observer or backward (B) tracing from observer towards
    the sun).}
  \label{tab:c2_models}
\end{table}
The settings of the Monte Carlo models that have run the C2 cases are summarized
in Table~\ref{tab:c2_models}. Except SPARTA all have used the same
number of photons (49\ee{9}) which corresponds to 10$^7$ photons per
pixel. 

The SHDOM resolution parameters were
$N_{\mu}$=32, $N_{\phi}$=64, $N_x$=140, $N_y$=140, $N_z$=101, and the cell splitting
accuracy was 0.001. The factor of two reduction in angular resolution
from case C1 was necessary due to the increased memory required by the
3D domain.  Since most of the domain volume is occupied by vacuum or
Rayleigh scattering, however, the SHDOM memory use is greatly reduced by
the adaptive spherical harmonics truncation.

\subsection{C2 -- Results}
\label{sec:c2_cubic_cloud_results}

We will discuss the results for three of the cases and show plots for all other
cases in \ref{sec:app}.

\begin{figure*}[htbp]
  \centering
  \includegraphics[width=1.\hsize]{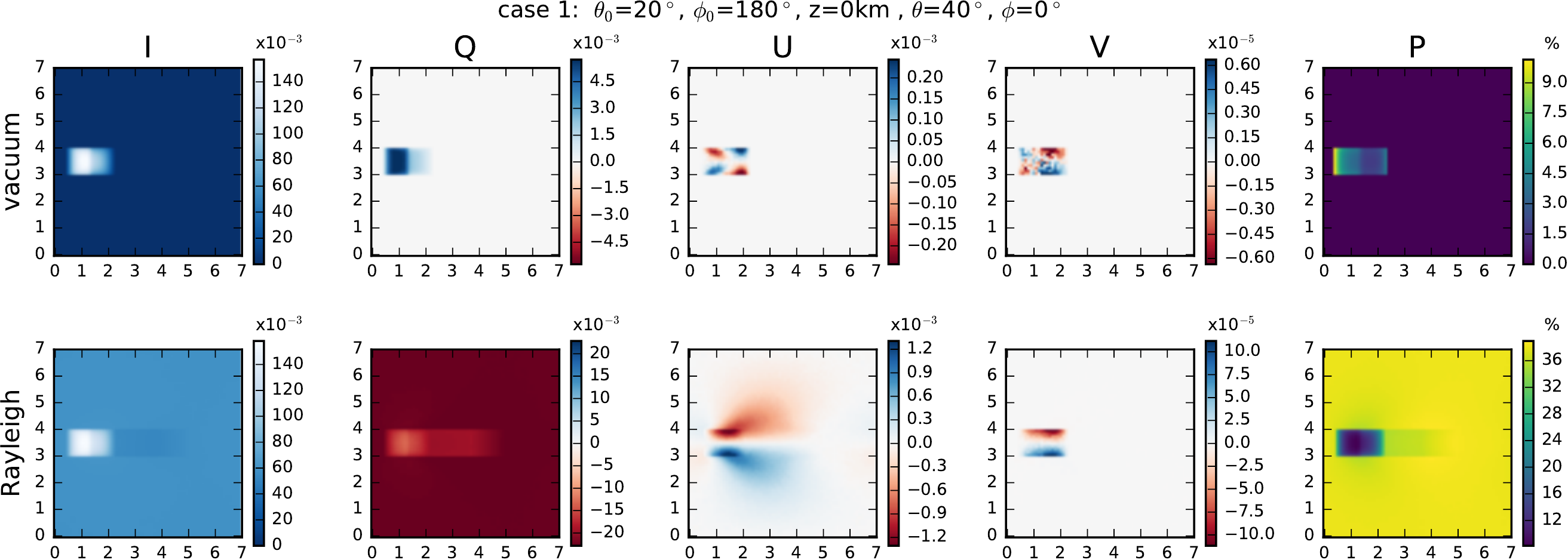}
  \caption{Results for scenario C2 (cubic cloud), case 1. The
    observer is at the ground and the viewing direction is
    $(\theta,\phi)=(40\degree,0\degree)$. The sun position is
    $(\theta_0,\phi_0)=(20\degree,180\degree)$. Upper panels: Cubic cloud
    is in vacuum. Lower panels: The cloud is embedded in a Rayleigh
    scattering layer. The labels on the x- and y-axes correspond to kilometers.}
  \label{fig:C2_subcase1}
\end{figure*}
Figure~\ref{fig:C2_subcase1} shows the MYSTIC results for case 1, where the
observer is at the bottom. The viewing direction is
$(\theta,\phi)=(40\degree,0\degree)$ and the sun position is
$(\theta_0,\phi_0)=(20\degree,180\degree)$. In the upper panels the cubic
cloud is in vacuum. The total radiance $I$ (upper left panel) shows that
the cloud is illuminated from the left side. When the cloud is
embedded in a Rayleigh scattering layer (lower panels) we see at the
right of the cloud its shadow in the atmosphere. The polarization
difference $Q$  
caused by pure cloud scattering is positive in this geometry with a value
of about 5\ee{-3} (radiance normalized to extraterrestrial irradiance
at 800~nm). A positive $Q$ means that the radiation is polarized parallel to the
scattering plane. 
The lower panel shows, that absolute value of $Q$ caused by Rayleigh
scattering is much larger (about 20\ee{-3}) and has a negative sign,
thus Rayleigh scattering polarizes perpendicular to the scattering
plane.  
The viewing direction is in the solar principal
plane, i.e., $U$ and $V$ would be exactly 0 for 1D (IPA)
radiative transfer
calculations due to the symmetry. The cubic cloud breakes the symmetry
and produces
characteristic patterns in $U$ and $V$. At the cloud edge, the degree
of polarization $P$ is up to 10\% for the cloud in vacuum. Rayleigh
scattering produces a $P$ of approximately 40\% which is reduced
inside the cloud and in the cloud shadow.  
\begin{figure*}[htbp]
  \centering
  \includegraphics[width=1.\hsize]{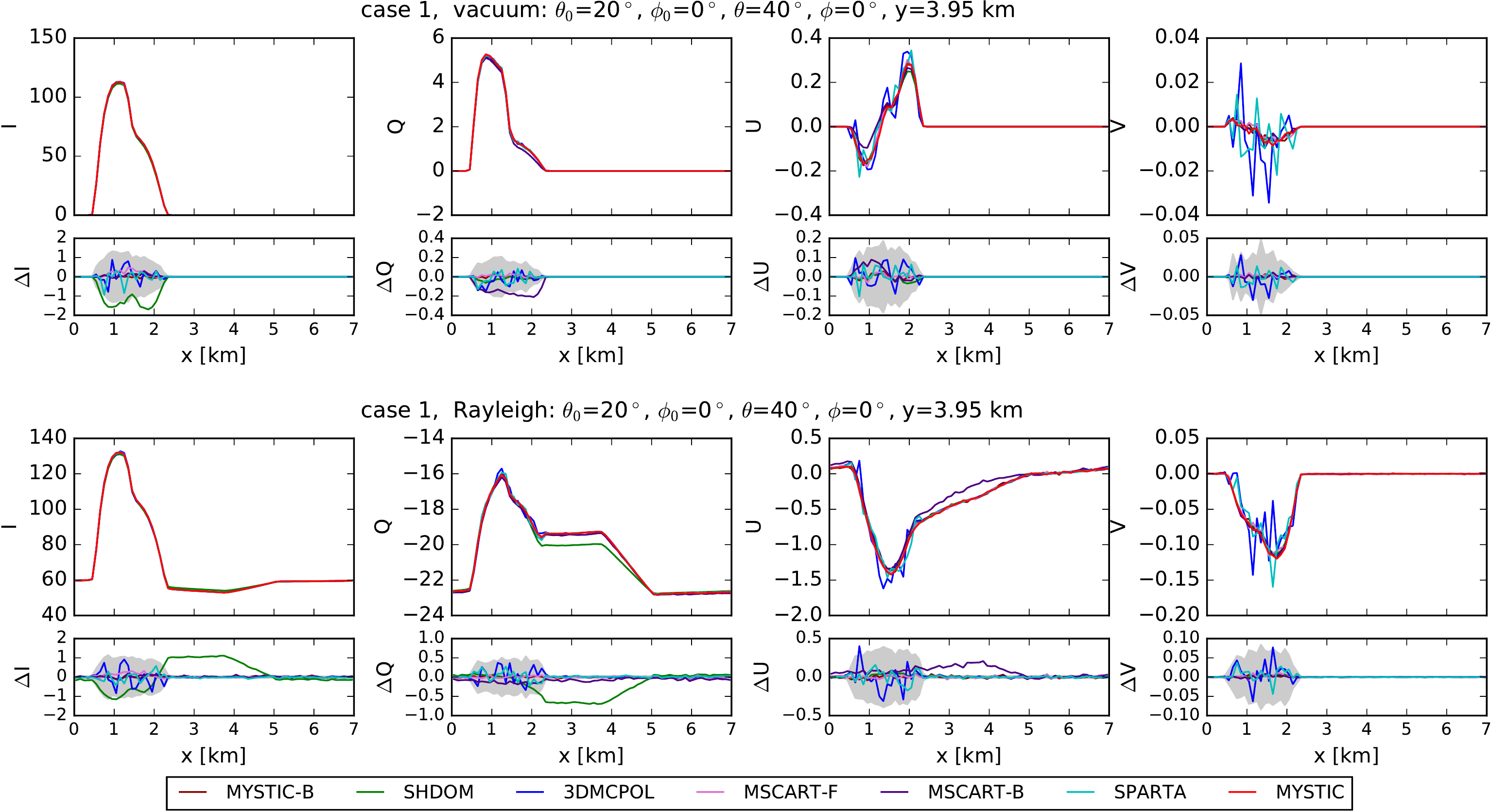}
  \caption{Results for scenario C2 (cubic cloud), case 1 for a cross
    section through the domain at y=3.95 km.  The upper panels are for
    the cloud in vacuum and the lower panels for the cloud embedded
    in a Rayleigh scattering layer. The plots include the values of
    the Stokes components $I$, $Q$, $U$ and $V$, and below the absolute
    differences  $\Delta I$, $\Delta Q$, $\Delta U$ and $\Delta V$ 
    between the individual model results and MYSTIC.
    The grey area in the difference plots corresponds to 2$\sigma$ of
    3DMCPOL
    (Monte Carlo model without variance reduction).
    The Stokes vector components are normalized to 1000/$E_0$.}
  \label{fig:C2_subcase1_diff}
\end{figure*}
The patterns which we see in Figure~\ref{fig:C2_subcase1} look the same
for all models. This is clearly seen in
Figure~\ref{fig:C2_subcase1_diff}, which shows all model
results for a cross section through the model domain at y=3.95~km. We
find quite large noise for Stokes components $U$ and $V$ for the
models 3DMCPOL and SPARTA, but we see that those results lie in the
range of the standard deviation of the 3DMCPOL model. 
We find small biases for SHDOM (green
lines) and for MSCART-B, those will be further analysed below.

\begin{figure*}[htbp]
  \centering
  \includegraphics[width=1.\hsize]{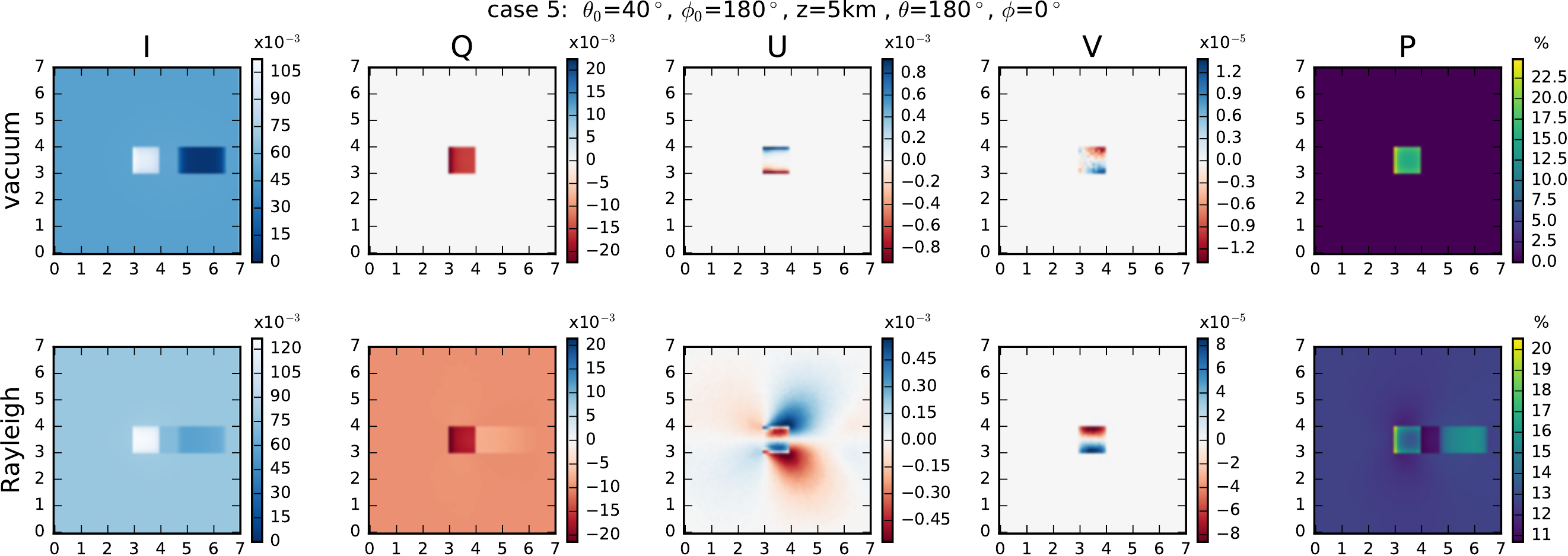}
  \caption{Results for scenario C2 (cubic cloud), case 5, for an
    observer at the top of the model atmosphere. The viewing direction is
    nadir and the sun position is
    $(\theta_0,\phi_0)=(40\degree,180\degree)$. Upper panels: Cubic cloud
    is in vacuum. Lower panels: The cloud is embedded in a Rayleigh
    scattering layer. The labels on the x- and y-axes correspond to kilometers.}
  \label{fig:C2_subcase5}
\end{figure*}
Figure~\ref{fig:C2_subcase5} shows the MYSTIC results for case 5, where the
observer is at the top of the atmosphere. The viewing direction is
nadir and the sun position is
$(\theta_0,\phi_0)=(40\degree,180\degree)$. The total radiance $I$
shows the cubic cloud in the center of the domain as seen from
above. Further, since the surface albedo is 0.2, we can clearly see
the cloud shadow at the ground, which is shifted to the right. Cloud
scattering and Rayleigh scattering both produce a negative $Q$ in this
geometry. Since the scattering angle is 140\degree\, the cloud
produces a high degree of polarization of more than 25\% and Rayleigh
scattering about 15\%. The degree of polarization in the cloud shadow
is larger than the Rayleigh background because $I$ is small in the
shadow.  
Again, we obtain patterns for $U$ and $V$,
which are exactly 0 in the principal plane for 1D RT simulations. 

\begin{figure*}[htbp]
  \centering
  \includegraphics[width=1.\hsize]{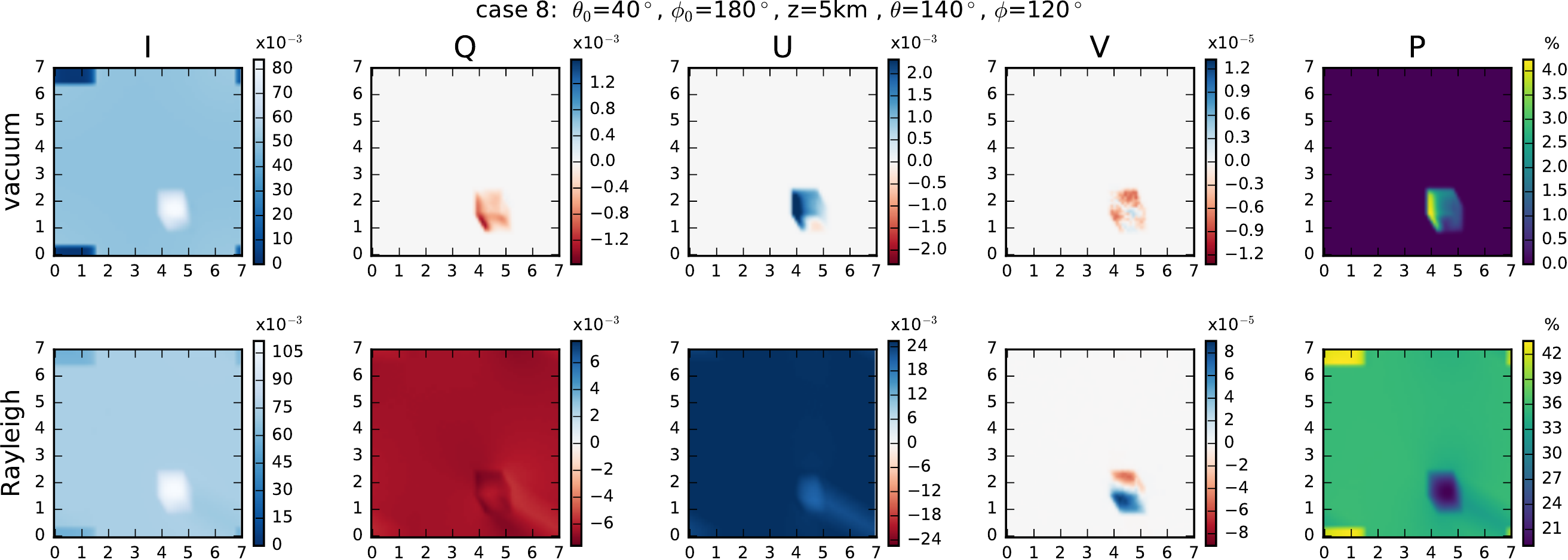}
  \caption{Results for scenario C2 (cubic cloud), case 8. 
    The sensor is at the top of the atmosphere and its viewing direction is
    $(\theta,\phi)=(140\degree,120\degree)$. The sun position is
    $(\theta_0,\phi_0)=(40\degree,180\degree)$. Upper panels: Cubic cloud
    is in vacuum. Lower panels: The cloud is embedded in a Rayleigh
    scattering layer. The labels on the x- and y-axes correspond to kilometers.}
  \label{fig:C2_subcase8}
\end{figure*}
MYSTIC results for case 8 are shown in Figure~\ref{fig:C2_subcase8}, where the
sun position is the same as for case 5. The sensor is at the top
of the atmosphere with viewing direction
$(\theta,\phi)=(140\degree,120\degree)$, i.e., outside the solar
principal plane. Here, we expect larger values for $U$ which are indeed
found in the results. For this geometry, $Q$ is positive for Rayleigh
scattering and cloud scattering and $U$ is mostly negative, with a
small positive area at the cloud edge. The degree of polarization is
about 35\% in the clear-sky region and about 20\% in the cloud
region. Due to periodic boundary conditions, the cloud shadow is
shifted into the upper and lower left corner of the domain. 
The degree of polarization in the cloud shadow is about 45\%.

\begin{figure*}[htbp]
  \centering
  \includegraphics[width=.9\hsize]{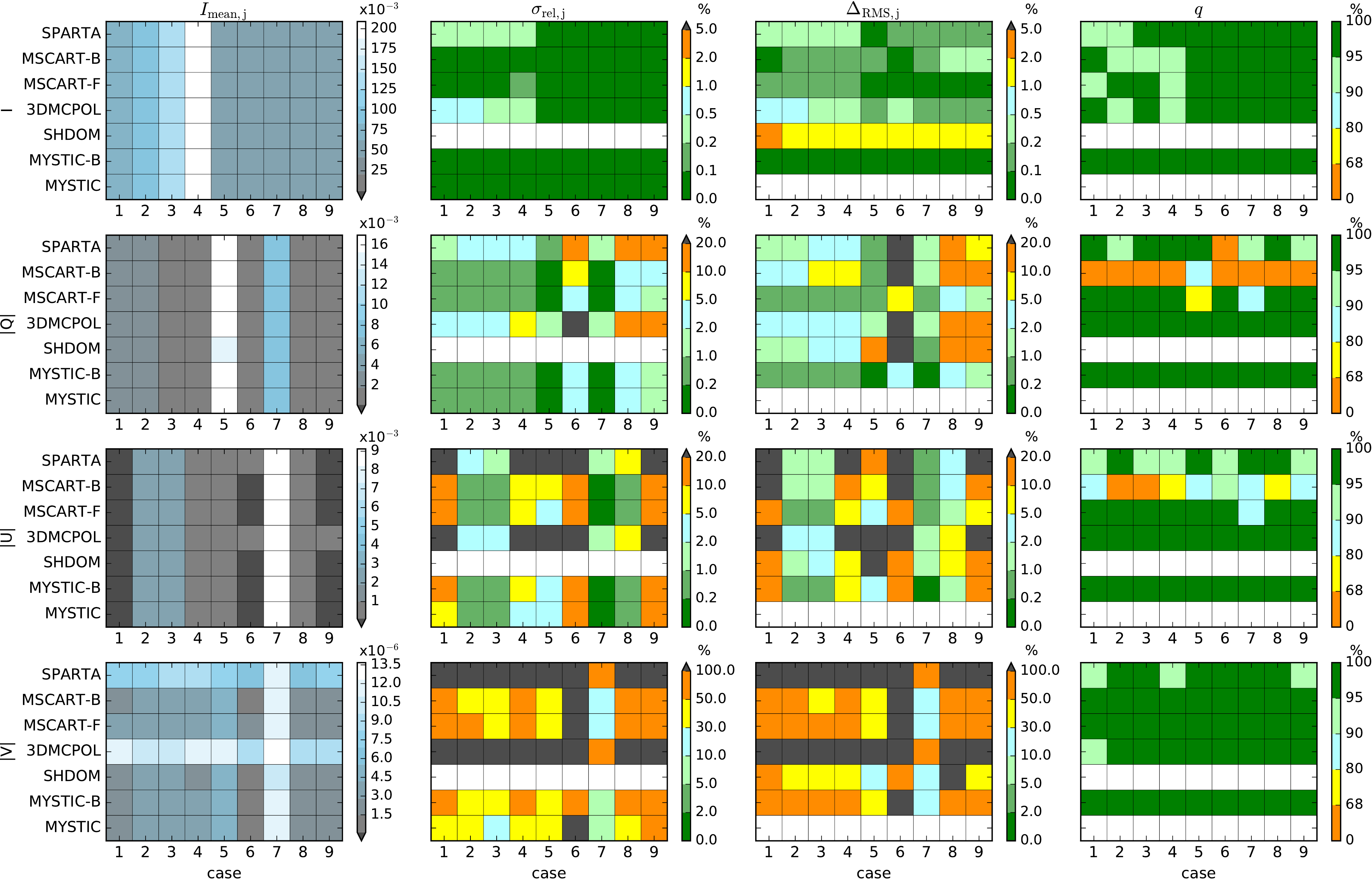}
  \caption{Statistics of the Stokes vector results for scenario C2
    (cubic cloud in vacuum).
    The panels in the left column show the mean radiance $I_{\rm mean}$ 
    (for Q, U, and V
    the mean of the absolute values) for all
    models and all 9 cases. The panels in the second column show the
    standard deviations $\sigma_{\rm rel}$. The third column shows the root mean
    square differences $\Delta_{\rm RMS}$ with respect to MYSTIC in per cent and the right
    column shows the match fractions $q$.}
  \label{fig:C2_stats_vacuum}
\end{figure*}
Figure~\ref{fig:C2_stats_vacuum} shows the statistics of the model
results for the cubic cloud in vacuum. All quantities were calculated
according to the definitions in Sec.~\ref{sec:stats}. The mean
radiance $I_{mean}$ (panels in left column) is very similar for all models and
all cases for the Stokes components $I$, $Q$ and $U$. 

For the circular polarization component
$V$, which is about three orders of magnitude smaller than $Q$ and $U$ we find
significant differences. In particular, \meanradV\ is larger for
SPARTA and 3DMCPOL compared to other models. The reason is that those two models have not used
variance reduction methods so that \relstdV\ is larger
than the absolute value of $V$.
The second column shows, that \relstdV\
is more than 100\% for SPARTA and 3DMCPOL for all
cases except case 7. Also, 
other models show very large noise for $V$; the only case with
\relstdV\ smaller than 30\% is case 7, where
the mean value of \meanradV\ is the largest.

For the first Stokes component (upper row in
Figure~\ref{fig:C2_stats_vacuum}),
\relstdI\ of all Monte Carlo codes is below 0.1\% for down-looking directions
(cases 5--9). For up-looking directions (cases 1--4) \relstdI\ is
slightly larger for the models SPARTA and 3DMCPOL but still below
1\%. 
The root mean square difference \rmsI\ is
similar to \relstdI\ for all cases and all Monte Carlo models. 
For SHDOM we obtain
\rmsI\ values of about 1--2\%, because SHDOM uses a different method
to solve the VRTE, in particular also a different grid
discretization.  The match fraction $q_I$ is mostly larger than 95\%,
i.e., the models agree within the Monte Carlo noise. For up-looking
cases, we often find values between 90\% and 95\%, which means that the
number of matches is slightly smaller than the expected assuming Gaussian
statistics. This can be explained by the small number of contributing
pixels  ($N_{\rm all}$\,$\approx$200),
because in the up-looking geometry only pixels within the cloud have
radiance values different from 0.
For SHDOM, we do not include a match
fraction because SHDOM does not calculate a standard deviation since
it applies a deterministic method.

For SHDOM, \rmsI\ is mostly about 1--2\%;  \rmsQ\ and \rmsU\ are in the
range from 1--20\%, with larger values for down-looking directions.
 
The plots of  \relstdQ\ and \relstdU\ show, 
that the variance reduction methods included
in MYSTIC and in MSCART work very well and reduce the relative
standard deviation from values between 2-5\% to values below 1\% for
a given number of photons. 
We find that MSCART-B (backward tracing mode)
disagrees to the other models, as the match fractions $q_Q$ and $q_U$ 
are almost always smaller than 90\%. The
main reason for this discrepancy arises from the invalidation of the
reciprocity principle when using a scattering-order-dependent phase
function truncation approximation \citep{wang2017,iwabuchi2009} in
backward tracing mode. For the backward simulation with this kind of
truncation approximation, the truncation fraction gradually increase
with decreasing scattering order, since the photons are traced from
detector (corresponding to the last order) to source (corresponding to
the first order). This means that, for the nth-order radiance
estimation, the greatest truncation occurs at the first-order
scattering simulation while the smallest one at the nth-order. 
The scattering-order-dependent phase
function truncation approximation produces a bias which is small for
forward tracing and can become larger in backward tracing mode.  

For MYSTIC, which aplies the variance reduction methods VROOM
\citep{buras2011}, we obtain the same results in forward and in
backward tracing mode (MYSTIC-B).

For several cases \relstdQ, \relstdU, \rmsQ\ and \rmsU\ are larger
than 20\%. These are the cases where the polarized radiance values are
very small.
The dark grey values in the plots for $I_Q$ and $I_U$ are values
below 10$^{-4}$, thus a relative standard deviation or relative root
mean square error is not meaningful for these cases. 
The mean radiance $I_V$ is always smaller than  10$^{-5}$, therefore
\relstdV\ and \rmsV\ are always large and a quantitative comparison
for circular polarization is not
possible. 

\begin{figure*}[htbp]
  \centering
  \includegraphics[width=.9\hsize]{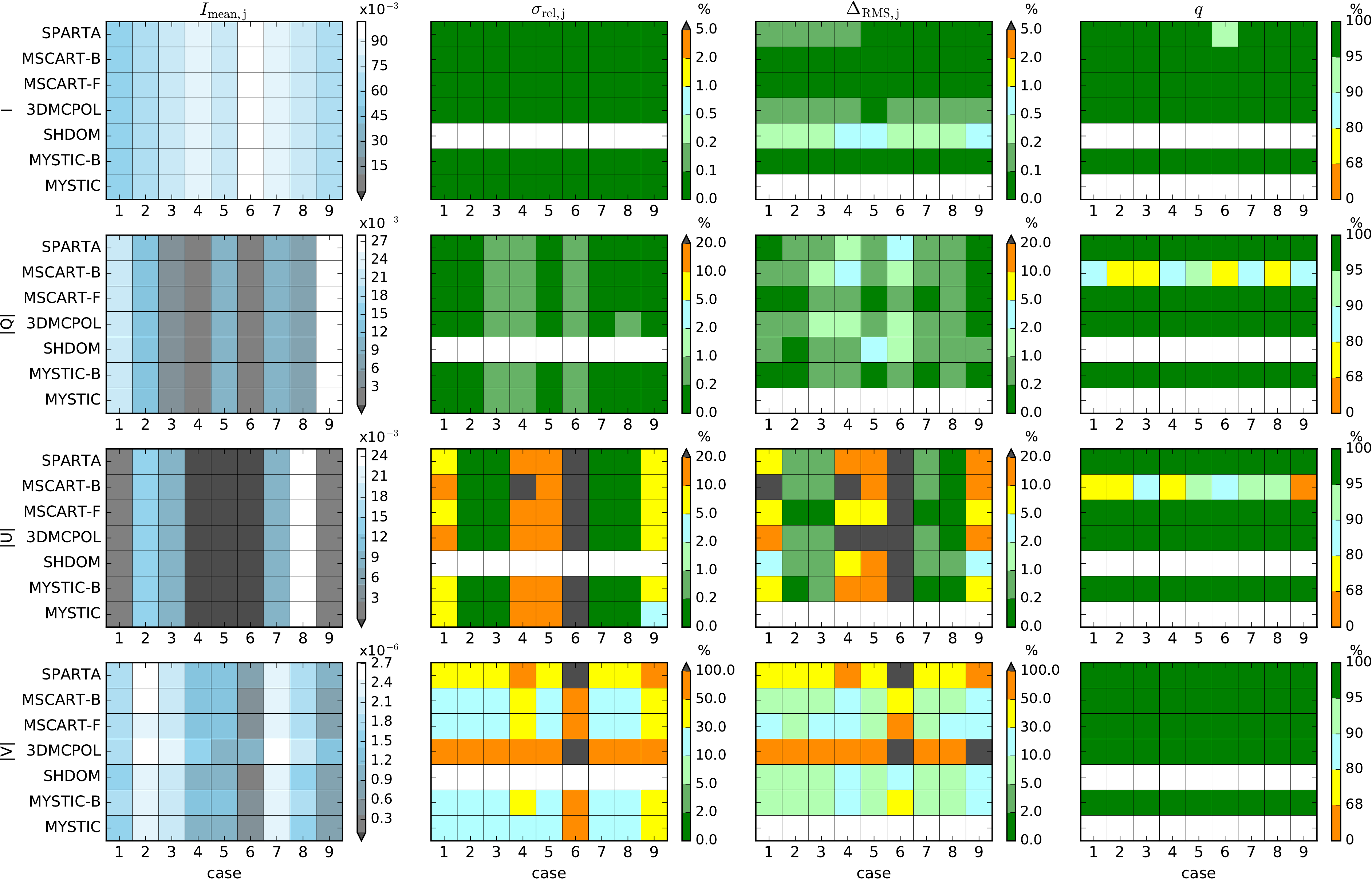}
  \caption{Statistics of the Stokes vector results for scenario C2
    (cubic cloud in Rayleigh layer).
    The panels in the left column show the mean radiance $I_{\rm
      mean}$ (for $Q$, $U$, and $V$
    the mean of the absolute values) for all
    models and all 9 cases. The panels in the second column show the
    standard deviations $\sigma_{\rm rel}$. The third column shows the root mean
    square differences $\Delta_{\rm RMS}$ with respect to MYSTIC in per cent and the right
    column shows the match fractions $q$. 
  }
  \label{fig:C2_stats_atm}
\end{figure*}
The statistics for the cubic cloud embedded in a Rayleigh scattering
layer is shown in Figure~\ref{fig:C2_stats_atm}. Generally, the results
become more accurate than for the cloud in vacuum, because Rayleigh
scattering is not characterized by strongly peaked
scattering phase functions as cloud scattering. Monte Carlo models can
accurately simulate Rayleigh scattering for relatively thin planetary
atmospheres such as the Earth's atmosphere
without variance reduction methods. However, \citet{garciamunoz2015}
show that for optically thick planetary atmospheres with strong
Rayleigh scattering ($\tau$=16) the directional sampling as
implemented e.g. in MYSTIC \citep{emde2010} causes numerical problems
so that the results do not converge.  To overcome this problem they
developed the ``pre-conditioned backward Monte Carlo'' method which
yields convergent results also for optically thick planetary
atmospheres.

The standard deviation \relstdI\ is generally smaller than 0.1\%,
\relstdQ\ and \relstdU\ are smaller than 1\% for all 
cases with mean values of $I_{\rm mean,j}$ larger than
approximately 3\ee{-3}. As for the cloud in vacuum, MSCART in backward
tracing mode yields significantly deviating results for $Q$ and $U$,
whereas in forward tracing mode it agrees perfectly to
the other codes. For SPARTA we find a significant difference for $I$
(case 6). For SHDOM \rmsI, \rmsQ, and \rmsU\ are smaller than
1\%.

\begin{figure*}[htbp]
  \centering
  \includegraphics[width=.9\hsize]{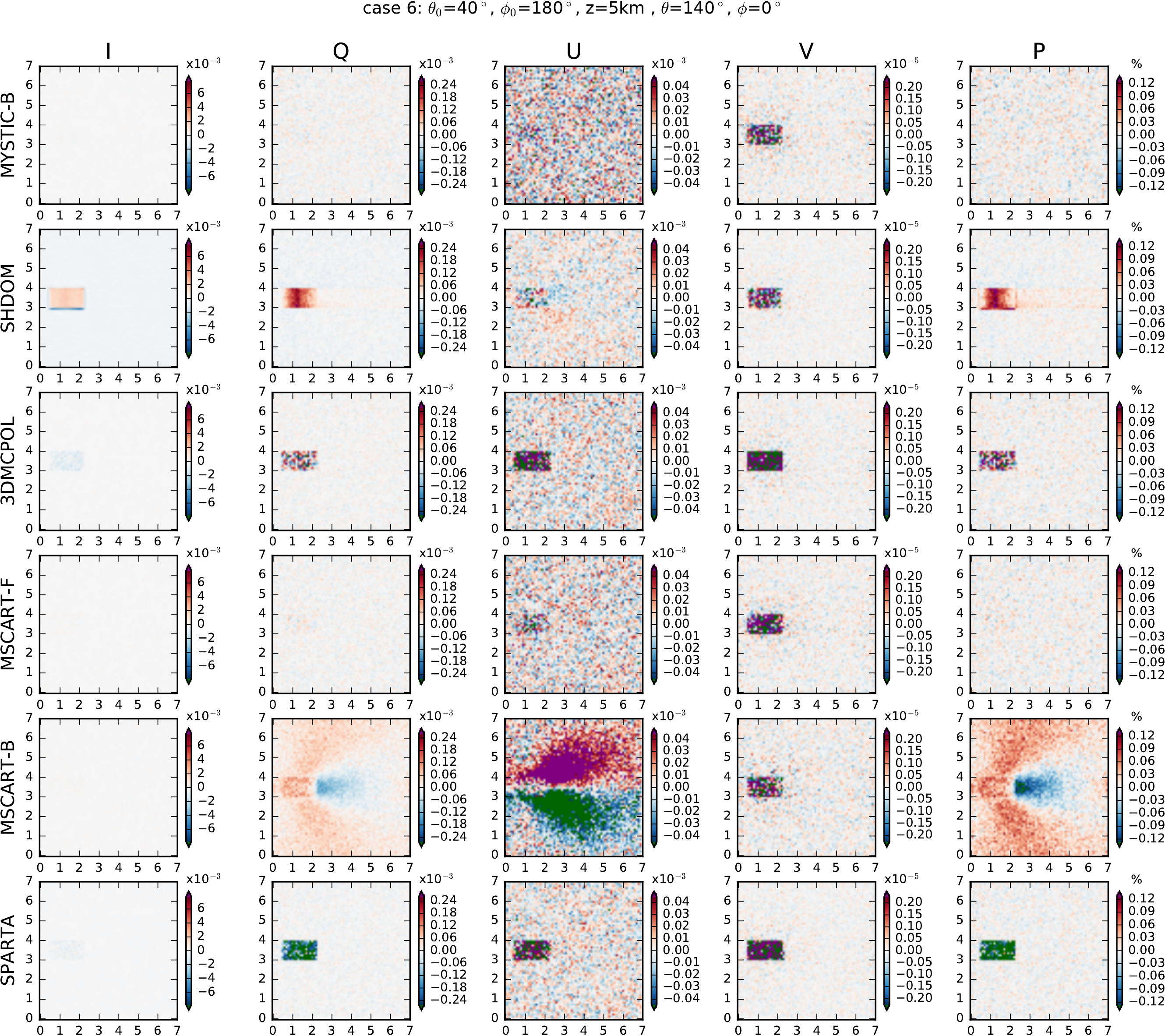}
  \caption{Absolute differences for Stokes vector components $I$, $Q$,
    $U$, $V$, and the degree of polarization $P$ between individual
    model results  and
    MYSTIC ($I_{i,model} - I_{i,MYSTIC}$)
    for cubic cloud in Rayleigh layer (C2), case 6.
    Each row corresponds
    to a different model, see labels on the left.  The labels on the x- and y-axes correspond to kilometers.
  }
  \label{fig:C2_diff_subcase6}
\end{figure*}
To further analyze model discrepancies, we show the absolute differences between
MYSTIC and individual model results for case 6
in Figure~\ref{fig:C2_diff_subcase6}. Case 6 (direct backscattering)
is the most problematic
case which shows discrepancies for SHDOM, SPARTA, and MSCART-B.
The limits of the colorbars in the figure are set to 5\% of
the maximum value of the MYSTIC results for $I$ and $P$, and to 10\%
of the MYSTIC results for $|Q|$ and $|U|$. Values larger than the upper
limit are marked in purple and values smaller than the lower limit
are marked in green. 

The differences between MYSTIC and MYSTIC-B show
only statistical noise according to the standard deviations of the
results, thus MYSTIC obtains the same results in forward tracing and
backward tracing mode, as expected. 

In direct backscattering geometry, SHDOM gives lower accuracy using
the TMS radiance method \citep{evans1998}
due to the finely structured glory peak of the cloud phase function.
The systematic differences between SHDOM and MYSTIC results are the
following: For $I$,
SHDOM obtains slightly smaller values than MYSTIC outside the cloud and
larger values inside the cloud. On the cloud boundary, SHDOM results
are also smaller than MYSTIC.  Inside the cloud, $Q$ obtained with SHDOM
is  larger than MYSTIC. The larger $Q$ value within the cloud 
results in a larger degree of polarization $P$; it is about 0.1\% larger for
SHDOM compared to MYSTIC. 
 
The difference between 3DMCPOL and MYSTIC exhibits a tiny negative bias
in the cloud area for $I$. All other Stokes components and also $P$
show only statistical noise and no systematic differences. 
MSCART-F agrees perfectly to MYSTIC, only statistical noise is
visible in the difference plots. 
MSCART-B shows obvious systematic differences for $U$ and $V$, and thus,
in $P$. These differences are probably due to the variance
reduction method in MSCART (see explanation above). 
For SPARTA there is a tiny negative bias in the cloud region for $I$
and a larger negative bias for $Q$, which becomes
visible also in the difference plot for $P$. 

In summary, we may conclude that, with a few exceptions, the models
MSCART-F, MYSTIC, MYSTIC-B, 3DMCPOL, and SPARTA, yield equal
results for the cubic cloud scenario in vacuum and in a Rayleigh scattering
atmosphere. ``Equal results'' means that the differences between the
models is smaller than two standard deviations for more than 96\% of
the calculated pixels.
SHDOM results have larger \rms\ than the Monte Carlo codes mainly due to the
inherently lower angular and spatial resolution of a deterministic
representation code.
MSCART-B produces significant biases in $Q$ and $U$ for the
setup including the cubic cloud.

\section{Test case C3 -- Cumulus cloud field}
\label{sec:c3_cumulus}

\subsection{C3 -- Model setup}
\label{sec:c3_cumulus_setup}

\begin{figure}[htbp]
  \centering
  \includegraphics[width=0.9\hsize]{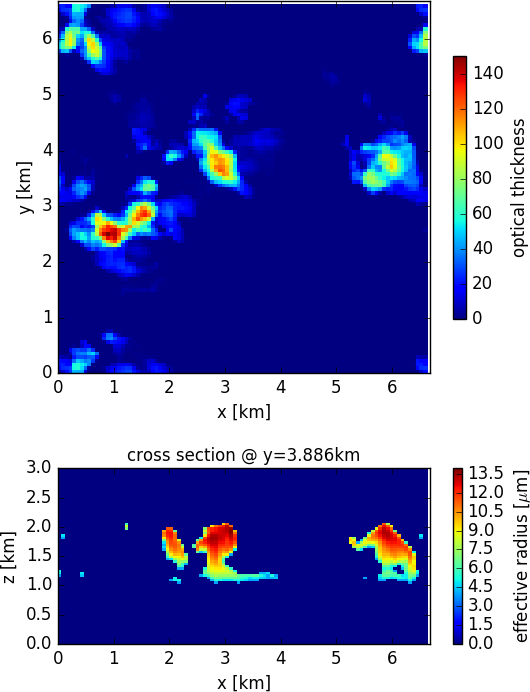}
  \caption{Definition of cumulus cloud, test case C3. }
  \label{fig:c3_cumulus_cloud_def}
\end{figure}
\begin{figure}[htbp]
  \centering
  \includegraphics[width=1.\hsize]{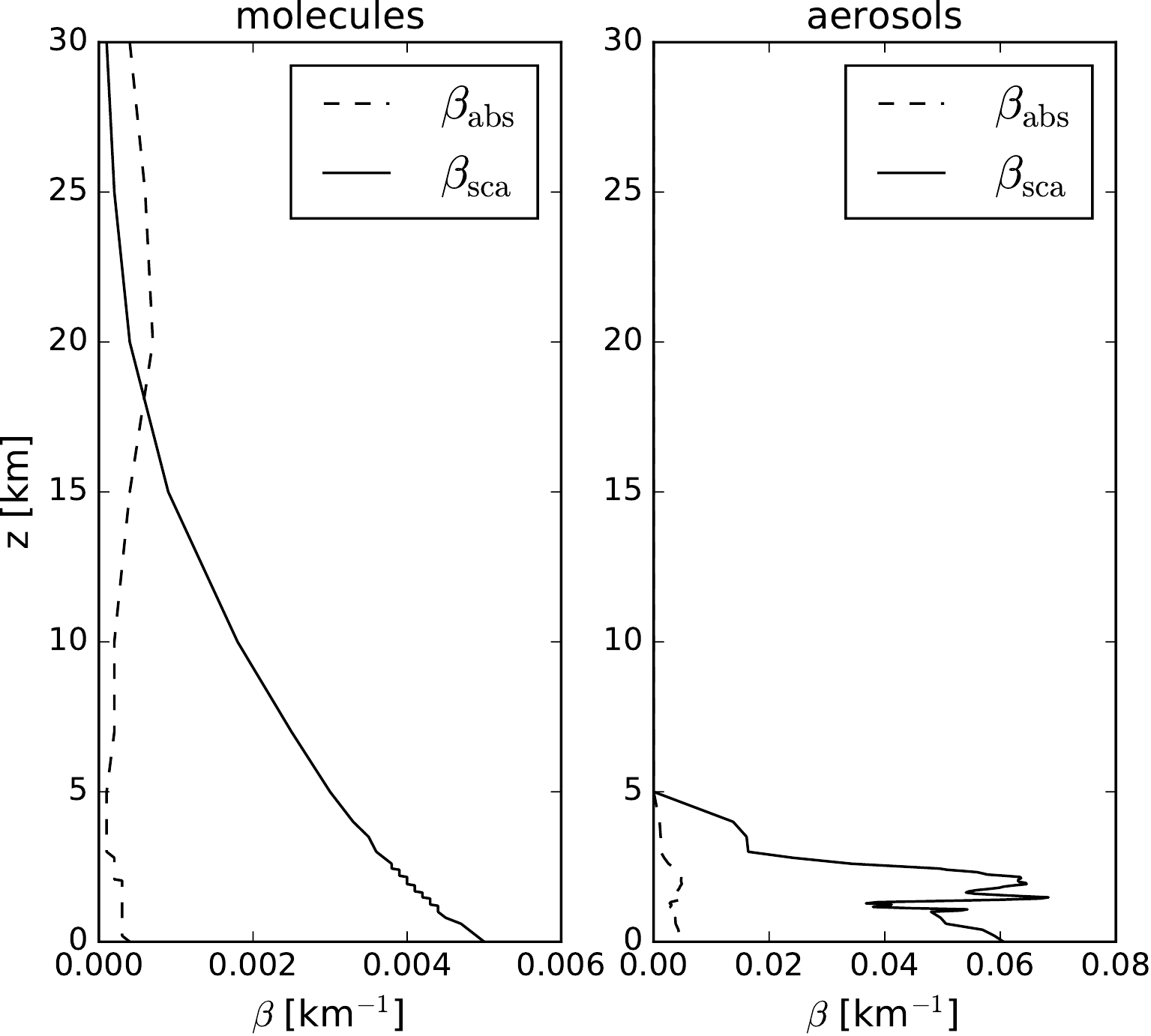}
  \caption{Altitude profiles of absorption coefficients $\beta_{\rm
      abs}$ and scattering coefficents $\beta_{\rm
      sca}$ coefficients for molecules (left) and aerosols (right).
    These profiles were used as atmospheric background for
    the simulations of the shallow cumulus field C3. 
  }
  \label{fig:C3_tau_profiles}
\end{figure}

This test case includes a shallow 
cumulus cloud field from large-eddy simulations (LES) by
\citet{stevens1999}, the same field was
also used in the I3RC (Intercomparison of 3D radiation Codes) project
\citep{cahalan2005}. The cloud field consists of
100$\times$100$\times$36 grid cells with a size of
66.7$\times$66.7$\times$40\,m$^3$. Extinction coefficients and
effective droplet radii for each grid cell were provided as model input data.
The upper panel in 
Figure~\ref{fig:c3_cumulus_cloud_def} shows the vertically integrated optical
thickness of the clouds and the lower panel shows the effective radius
for a vertical cross section at y\,=\,3.886\,km.
As for the cubic cloud case (C2), two sets of simulations were
performed at a wavelength of 670\,nm. 
In the first set of simulations, the cloud field is embedded
in a molecular atmosphere; altitude profiles of 
molecular absorption coefficients and
Rayleigh scattering coefficients were given as input for all models. 
The Rayleigh depolarization factor was set to 0.
The second set of simulations additionally includes aerosols;
the altitude profile of aerosol extinction coefficients were also
provided as model input. Figure~\ref{fig:C3_tau_profiles} shows altitude
profiles of absorption and scattering coefficients of molecules
and aerosols. 
Aerosol and cloud optical properties were
precalculated using Mie theory and provided as model input.
For aerosols, we used the
refractive index and size distribution parameters for water soluble
aerosol from the OPAC database \citep{hess1998}. The single scattering
albedo is approximately 0.93. 
A Lambertian surface was included with albedo 0.2 and, as for scenario C2,
the solar azimuth angle is 180\degree\ in all test cases.
Output altitudes, solar zenith
angles and viewing directions are the same as for scenario C2 (see
Table~\ref{tab:c2_settings}). 

\begin{table}
  \centering
  \begin{tabular}{l l l l l}
    model & $N_{ph}$ & VR & TM \\ \hline 
    3DMCPOL & 10$^{10}$ & no & F \\
    SPARTA & 10$^{10}$ & no & F \\
    MSCART-F & 10$^{10}$ & yes & F \\
    MSCART-B & 10$^{10}$ & yes & B \\
    MYSTIC & 10$^{10}$ & yes & F \\
    \hline 
  \end{tabular}
  \caption{Monte Carlo model settings for scenario C3 -- cumulus cloud
    field.
    $N_{ph}$ is the number of photons, the VR column shows whether
    the models used variance reduction methods for spiky scattering phase
    functions and TM gives the tracing method (forward tracing (F) from
    the sun towards observer or backward (B) tracing from observer towards
    the sun).}
  \label{tab:c3_models}
\end{table}
Table~\ref{tab:c3_models} shows the settings of the Monte Carlo models
that have run the C3 test cases with the cumulus cloud field. For
these test cases all models have run the commonly agreed number of photons of
10$^{10}$ which corresponds to 10$^{6}$ photons per pixel. Since the
number of photons per pixel is a factor of 10 smaller than for test cases C2, we
expect less accurate results (the standard deviation is expected to
increase by a factor of $\sqrt{10}$).  
The SHDOM resolution parameters were
$N_{\mu}=16$, $N_{\phi}$=32, $N_x$=100, $N_y$=100, $N_z$=55, and cell splitting
accuracy of 0.03.  The further reduction of the angular and spatial
resolution from case C2 was required by the large domain filled with
aerosols, for which the adaptive spherical harmonics truncation does not
save significant memory.

\subsection{C3 -- Results}
\label{sec:c3_cumulus_results}

In this section we will discuss the results for three of the cases,
plots of all other cases are shown in \ref{sec:app}.

\begin{figure*}[htbp]
  \centering
  \includegraphics[width=1.\hsize]{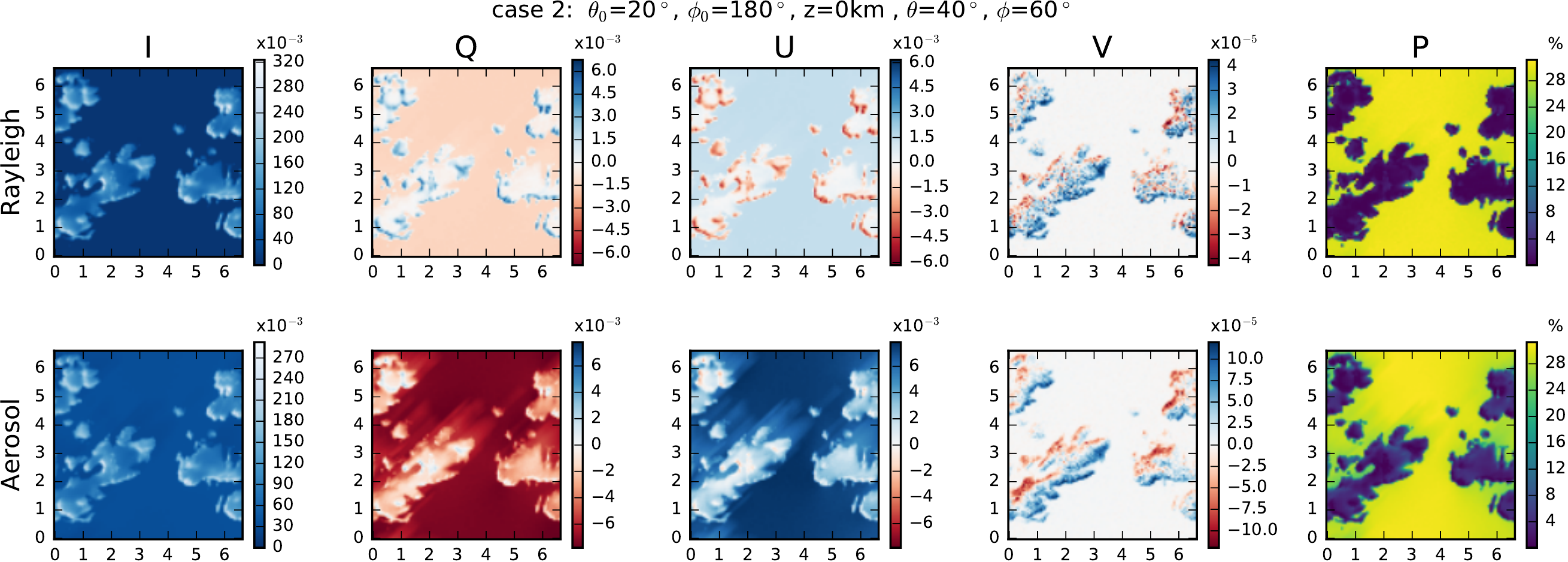}
  \caption{Results for scenario C3 (shallow cumulus cloud field), case 2, for an
    observer at the surface. The viewing direction is $(\theta,\phi)=(40\degree,60\degree)$
    and the sun position is
    $(\theta_0,\phi_0)=(20\degree,180\degree)$.
    Upper panels: The cumulus cloud field is surrounded by a molecular atmosphere.
    Lower panels: Aerosol is added to the scenario. The labels on the x- and y-axes correspond to kilometers. }
  \label{fig:C3_subcase2}
\end{figure*}
Figure~\ref{fig:C3_subcase2} includes the MYSTIC results for case 2, where the
observer is placed at the surface, viewing upwards into direction
$(\theta,\phi)=(40\degree,60\degree)$. The sun position is
$(\theta_0,\phi_0)=(20\degree,180\degree)$. The plot nicely shows that
the clouds are illuminated from the bottom-left side in this
geometry. The upper plots are for the clouds embedded in a pure
molecular atmosphere. $Q$ is negative in the clear-sky region and $U$
is positive.  Inside the clouds,  $Q$ and $U$
have opposite signs or the
polarization becomes close to 0. For the circular polarization
$V$, we see clear patterns with
negative values on the left hand side of the sun direction and positive
values on the right hand side. The degree of polarization is about 30\% in
the clear sky region and very small inside the clouds. 

The bottom plots are for the same molecular atmosphere, but including
aerosol in addition. We see in the $Q$ and $U$ plots that aerosol
produces the same signs as Rayleigh scattering, thus it enhances the
polarized radiance. The pattern in $V$ remains the same as without
aerosol, but the magnitude is larger and the Monte Carlo noise becomes
smaller. The degree of polarization is similar with and without
aerosol, because as $Q$, $U$, and $V$ are increased by aerosol
scattering, also the intensity $I$ is increased.  

\begin{figure*}[htbp]
  \centering
  \includegraphics[width=1.\hsize]{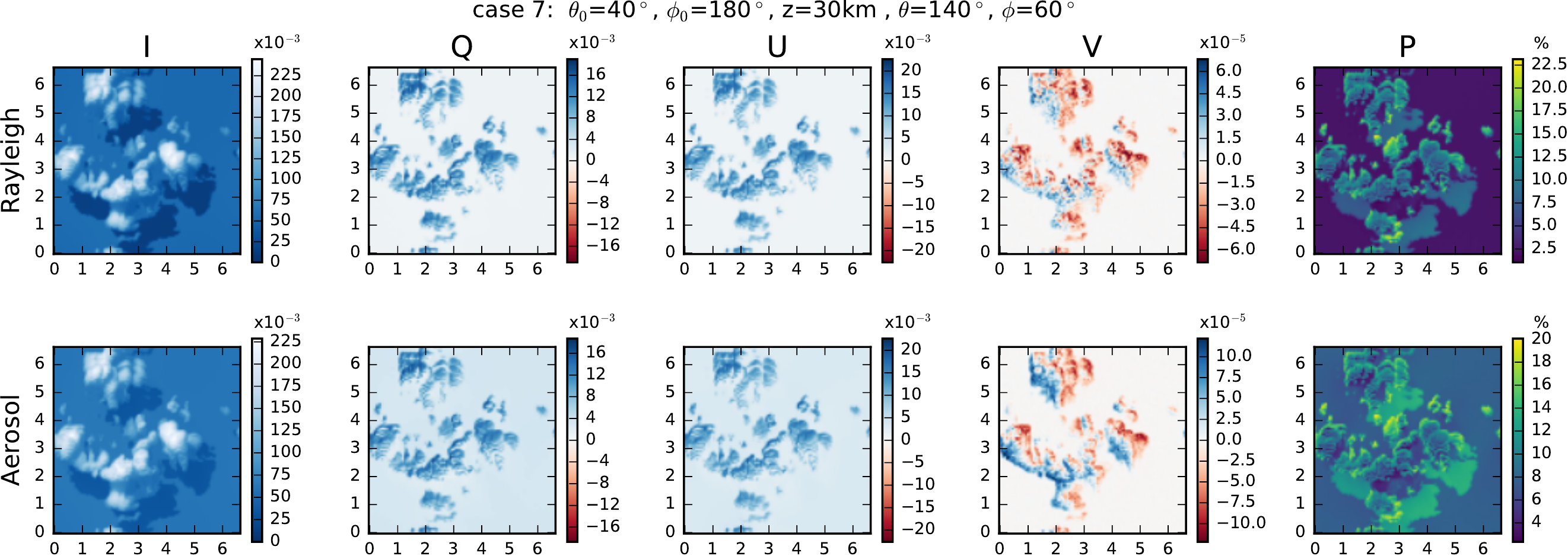}
  \caption{Results for scenario C3 (shallow cumulus cloud field), case 7, for an
    observer at the top of the model atmosphere. The viewing direction is
    $(\theta,\phi)=(140\degree,60\degree)$ and the sun position is
    $(\theta_0,\phi_0)=(40\degree,180\degree)$. Upper panels: The
    cumulus cloud field is surrounded by a molecular atmosphere.
    Lower panels: Aerosol is added to the scenario.The labels on the x- and y-axes correspond to kilometers.  }
  \label{fig:C3_subcase7}
\end{figure*}
Figure~\ref{fig:C3_subcase7} shows the MYSTIC results for case 7. Here, the
observer is at the top of the atmosphere and looks into direction
$(\theta,\phi)=(140\degree,60\degree)$. The sun direction is
$(\theta_0,\phi_0)=(40\degree,180\degree)$. Figure~\ref{fig:c2_geometry}
shows that in this geometry the scattering angle is 140\degree,
therefore we expect that the clouds generate polarization because we
know that the cloud-bow is polarized. The $Q$ and $U$ plots nicely show
the polarization by cloud scattering. $Q$ and $U$ are both positive
and the values inside the cloudy region are much larger than the
background. For aerosol, the background polarization is slightly
larger, but still it is much smaller than the polarization by cloud
scattering. For $V$, we see again symmetric patterns about the sun
direction. The degree of polarization in the clouds is roughly between
10\% and 20\%, where the highest values are obtained for thin clouds,
e.g., the small cloud at around $x$\,=6.2km and $y$\,=4.2km. In the image of the
radiance $I$, it appears relatively faint, whereas in the image of the degree of
polarization $P$ it is one of the brightest spots. The degree of
polarization in the thick clouds becomes very small due to multiple
scattering, which increases $I$, but not $Q$, $U$, and $V$ by the same
magnitude, because  $Q$, $U$ and $V$ saturate after a few scattering orders. 
The degree of polarization in the cloud shadows is
increased because $I$ is decreased. 
\begin{figure*}[htbp]
  \centering
  \includegraphics[width=1.\hsize]{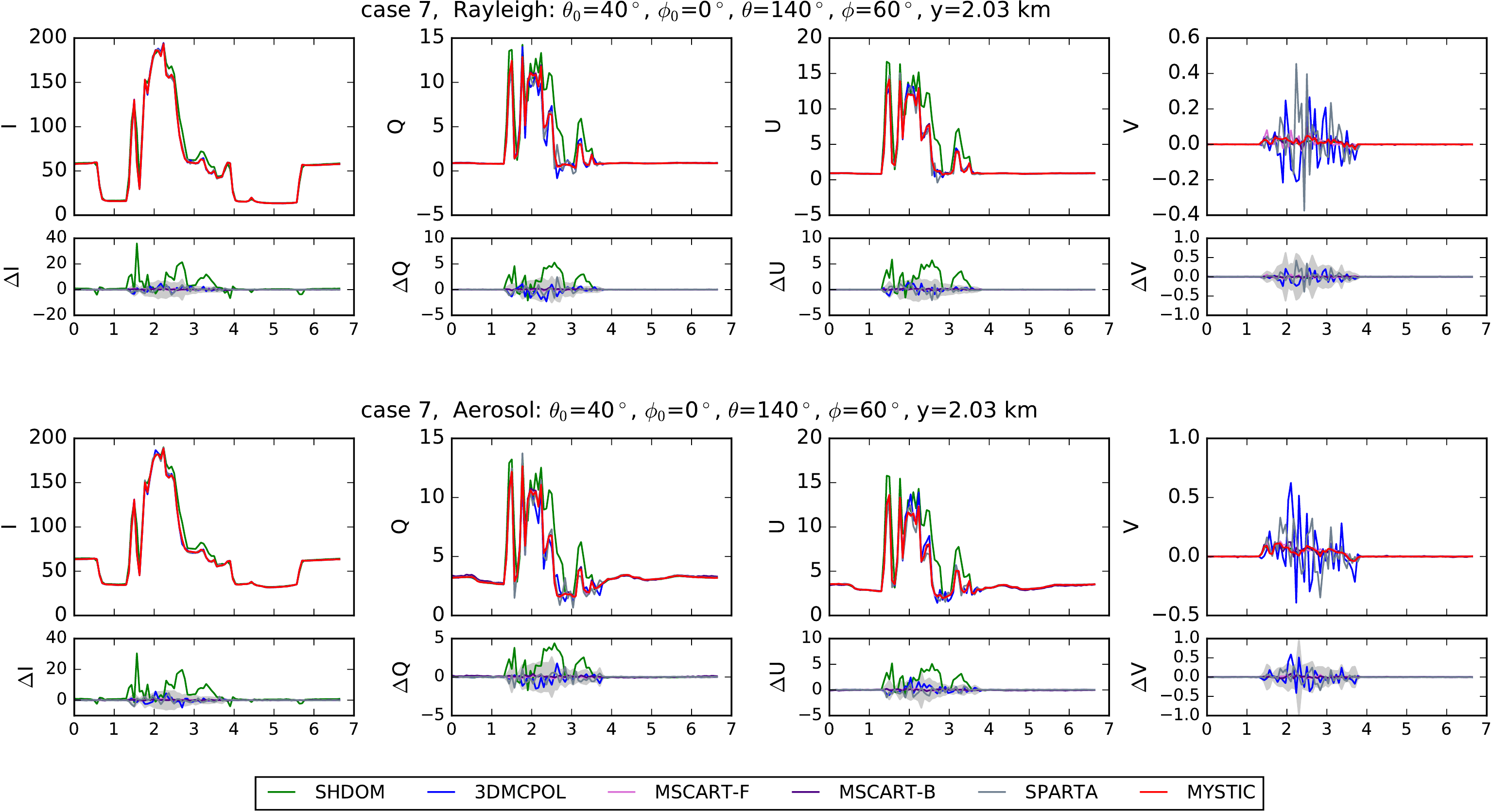}
  \caption{Results for scenario C3 (shallow cumulus cloud field), case
    7, for a cross
    section through the domain at y=2.03 km.  The upper panels are for
    the cloud in vacuum and the lower panels for the cloud embedded
    in a Rayleigh scattering layer. The plots include the values of
    the Stokes components $I$, $Q$, $U$ and $V$, and below the absolute
    differences $\Delta I$, $\Delta Q$, $\Delta U$ and $\Delta V$
    between the individual model results and MYSTIC.
    The grey area in the difference plots corresponds to 2$\sigma$ of
    3DMCPOL (Monte Carlo model without variance reduction).
    The Stokes vector components are normalized to 1000/$E_0$.}
  \label{fig:C3_subcase7_diff}
\end{figure*}
Figure~\ref{fig:C3_subcase7_diff} shows the results of all models for
a cross section though the model domain at y=2.03~km. For all Monte
Carlo codes, the results are very similar for $I$, $Q$, and $U$. For
$V$ the noise is much larger than the value of $V$ for models without
variance reduction methods (SPARTA, 3DMCPOL). For all Stokes
components differences between the Monte Carlo models and MYSTIC are
within the grey area defined by two standard deviations of 3DMCPOL. 
SHDOM results show the same spatial patterns but are slightly
biased against the Monte
Carlo results. This will be further discussed below. 

\begin{figure*}[htbp]
  \centering
  \includegraphics[width=1.\hsize]{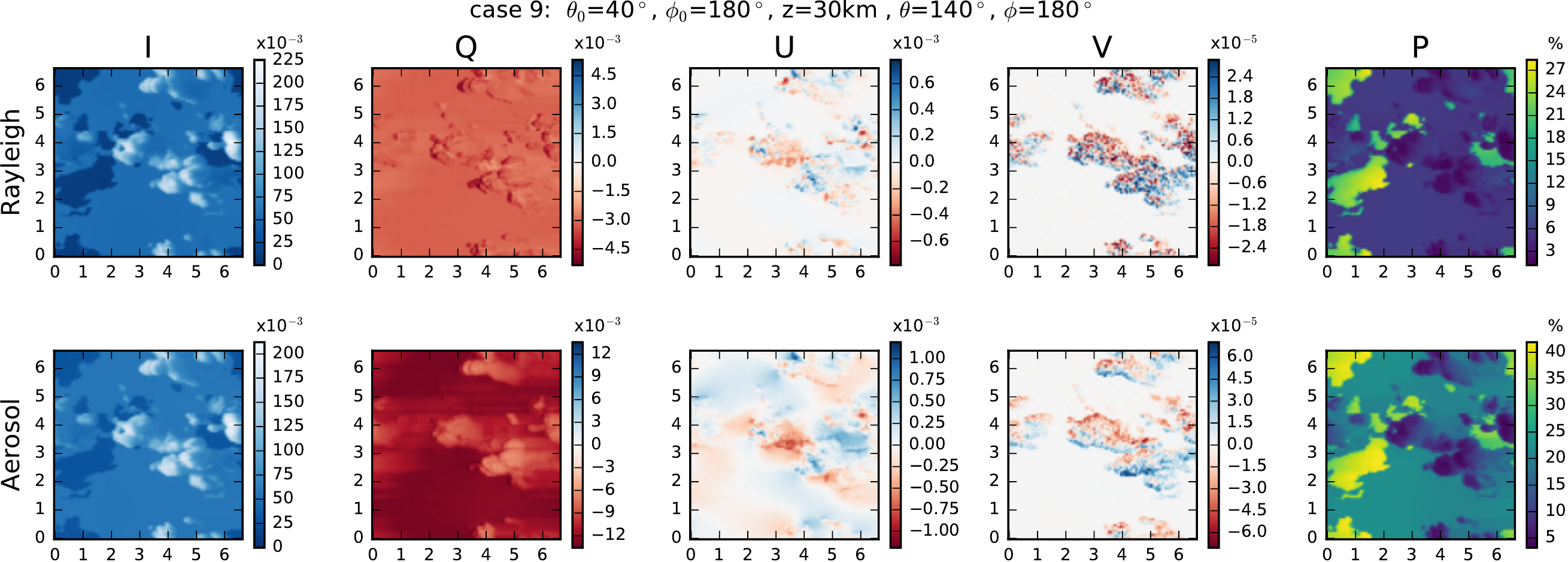}
  \caption{Results for scenario C3 (shallow cumulus cloud field), case 9, for an
    observer at the top of the model atmosphere. The viewing direction is
    $(\theta,\phi)=(140\degree,180\degree)$ and the sun position is
    $(\theta_0,\phi_0)=(40\degree,180\degree)$. Upper panels: The
    cumulus cloud field is surrounded by a molecular atmosphere.
    Lower panels: Aerosol is added to the scenario. The labels on the x- and y-axes correspond to kilometers.}
  \label{fig:C3_subcase9}
\end{figure*}
Figure~\ref{fig:C3_subcase9} shows the MYSTIC results for case 9. The
observer is placed at the top of the atmosphere and its viewing
direction is $(\theta,\phi)=(140\degree,180\degree)$. The sun position
is $(\theta_0,\phi_0)=(40\degree,180\degree)$. In this geometry with a
scattering angle of 80\degree, we
expect polarization from Rayleigh and aerosol scattering. Since we are in the
solar principal plane, the mean of $U$ and $V$ should be close to 0.
We obtain negative values of $Q$ from molecular scattering. Also,
clouds produce negative $Q$ polarization in this geometry, so the
clouds show higher absolute $Q$-values than the background in
particular on the side facing the sun. The degree of
polarization is about 5\% in the clear-sky region. When aerosol
particles are
added, the background becomes much more  polarized with negative
$Q$-values. The degree of polarization increases to about 25\%. The
absolute value of $Q$ is
now larger in the clear-sky region with aerosols than within the
clouds.

\begin{figure*}[htbp]
  \centering
  \includegraphics[width=.9\hsize]{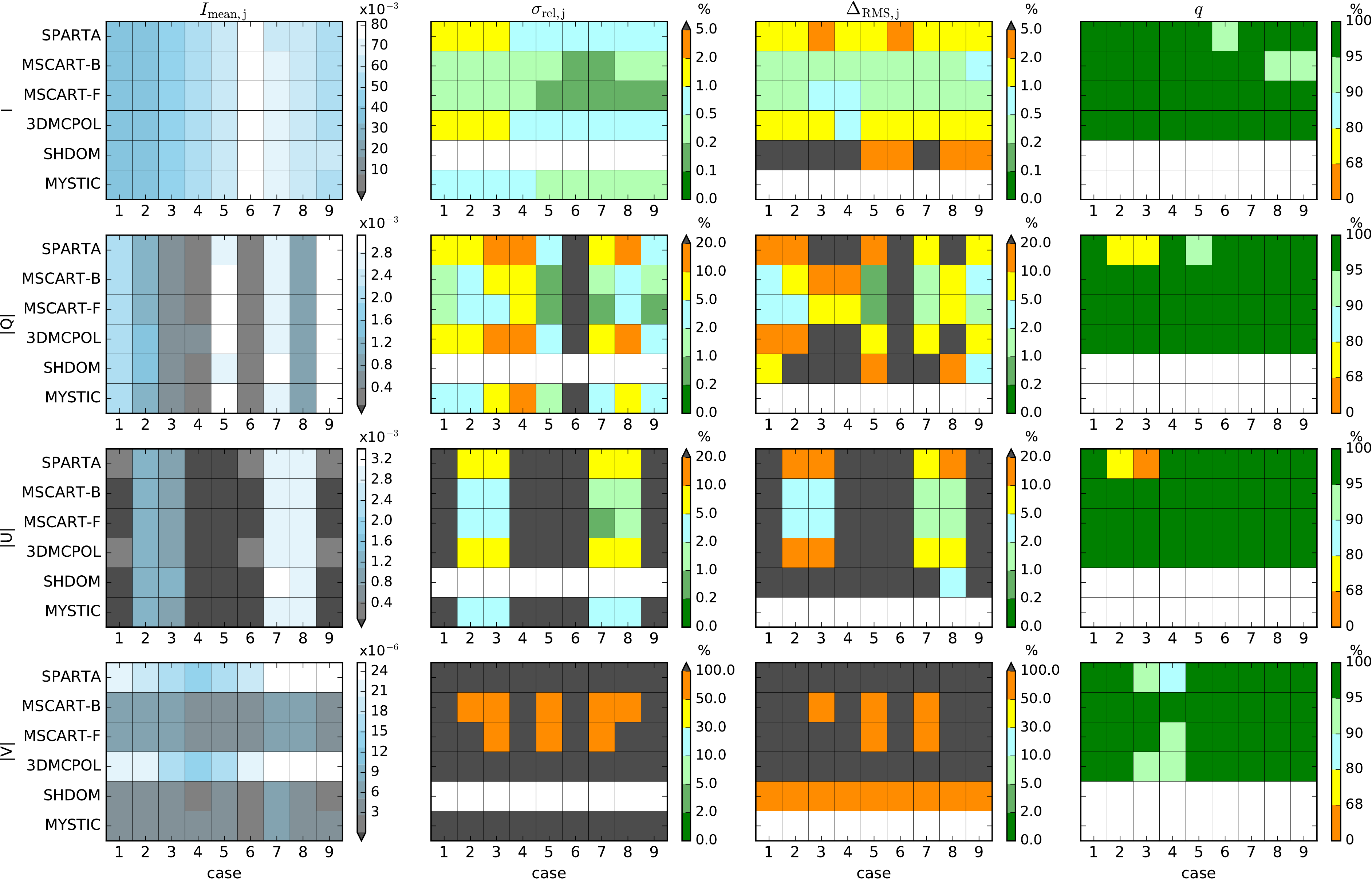}
  \caption{Statistics of the Stokes vector results for scenario C3
    (cumulus cloud field in molecular atmosphere).
    The panels in the left column show the mean radiance $I_{\rm
      mean}$ (for Q, U, and V
    the mean of the absolute values) for all
    models and all 9 cases. The panels in the second column show the
    standard deviations $\sigma_{\rm rel}$. The third column shows the root mean
    square differences $\Delta_{\rm RMS}$ in per cent and the right
    column shows the match fractions $q$. 
  }
  \label{fig:C3_stats_noaer}
\end{figure*}
Figure~\ref{fig:C3_stats_noaer} shows the statistics for the cumulus
cloud field in the molecular atmosphere. The mean radiance \meanradI\ is very
similar for all models. The mean values \meanradQ, \meanradU, and \meanradV\ show
several differences:
As for the cubic cloud scenario C2
(Section~\ref{sec:c2_cubic_cloud_results}), \meanradV\ is
significantly larger for 
SPARTA and 3DMCPOL than for the other models, because the
accuracy is lower for models without variance
reduction techniques. 
The mean values \meanradU\ are smaller than 10$^{-4}$ for
cases 1,4,5,6, and 9 (dark grey). For all these cases, \relstdU,
\rmsU, and $q_U$ are not meaningful. For case 6, the same applies for
\meanradQ.
  
The values of \relstdI\ are always below
2\% for SPARTA and 3DMCPOL.  For MSCART-F and MSCART-B, \relstdI\ is
below 0.5\%.
For MYSTIC \relstdI\ is smaller than 0.5\% for down-looking cases
5--9 and between 0.5\% and 1\% for up-looking directions.
This demonstrates once more that the
variance reduction method VROOM reduces the noise
better for down-looking than for 
up-looking directions.

The relative standard deviations \relstdQ\ and \relstdU\ are
below 5\% for models with variance reduction and cases where the
\meanradQ\ and \meanradU\ are larger than approximately 10$^{-3}$. 

The match fraction $q$ is larger than 95\% for most cases. 
Obvious differences are only found for cases 2, 3, and 5 for the SPARTA
model, where $q_Q$ and $q_U$ are smaller than 95\%.
Further for cases 8 and 9, $q_I$ is smaller than 95\% for MSCART-B;  
for case 6,  $q_I$ is smaller than 95\% for SPARTA.  

\begin{figure*}[htbp]
  \centering
  \includegraphics[width=.9\hsize]{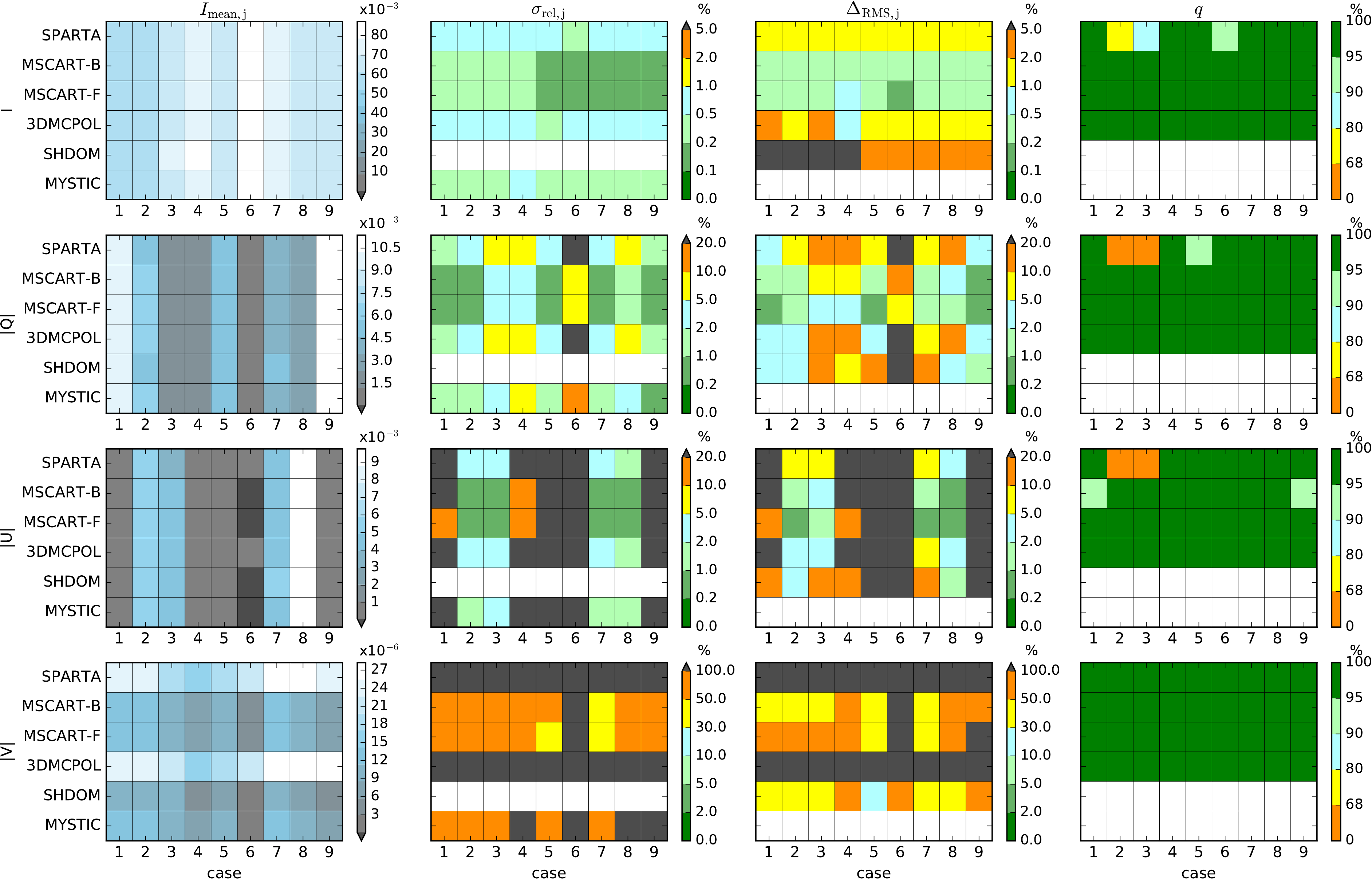}
  \caption{Statistics of the Stokes vector results for scenario C3
    (cumulus cloud field and aerosol in molecular atmosphere).
    The panels in the left column show the mean radiance $I_{\rm
      mean}$ (for Q, U, and V
    the mean of the absolute values) for all
    models and all 9 cases. The panels in the second column show the
    standard deviations $\sigma_{\rm rel}$. The third column shows the root mean
    square differences $\Delta_{\rm RMS}$ in per cent and the right
    column shows the match fractions $q$. 
  }
  \label{fig:C3_stats_aer}
\end{figure*}
Figure~\ref{fig:C3_stats_aer} shows the statistics for the cumulus
cloud field in the molecular atmosphere containing aerosols.
The mean values of the Stokes vector (left column) are similar
for all models. 

The relative standard
deviations \relstdI\ are below 0.5\% for MSCART-B, MSCART-F, and
MYSTIC. Only for case 4, \relstdI\ for MYSTIC is 
between 0.5\% and 1\%.
MSCART-F and MSCART-B yields the smallest values (\relstdI$<$0.2\%)
for the down-looking cases 5--9 which demonstrates that for
cases with aerosol and clouds, the variance reduction method of MSCART
is more efficient than VROOM. Also, \relstdQ\ and \relstdU\ are
often below 1\% for MSCART-F and MSCART-B, for MYSTIC the values are 
between 1\% and 2\%. The accuracy of SPARTA and 3DMCPOL is
lower, \relstdI\ is mostly between 0.5\% and 1\%; 
\relstdQ\ and \relstdU\ are in the range from 2\% to 5\% for mean values of
\meanradQ\ and \meanradU\ larger than about 10$^{-3}$. 

The \rms\ values are generally larger than \relstd\, because they
include the inaccuracies of the two models that are compared. 
For SHDOM, \rmsI\ is larger than for all Monte Carlo codes with values
between 2\% and 5\% for all down-looking cases 5--9, and more than 5\%
for the up-looking cases. For $Q$ and $U$, the \rms-differences of
SHDOM are comparable to those of SPARTA and 3DMCPOL. 

For circular polarization,
\relstdV\ and \rmsV\ are generally very high. Values below
100\% are obtained only for MSCART-F, MSCART-B, and MYSTIC. The
\rmsV-difference between MSCART-F, MSCART-B, and MYSTIC is comparable to
the \rmsV-difference between SHDOM and MYSTIC. 

The match fraction $q$ reveals problems for SPARTA, mainly for test
cases 2 and 3.  Smaller differences for
SPARTA also exist for case 5 (see $q_Q$) and case 6 (see $q_I$). 
Further we again see a small deviation for MSCART-B for cases 1 and 9
for the $U$-component. However, for these cases the mean value of \meanradU\
is smaller than 10$^{-4}$, thus these differences are not meaningful. 

\begin{figure*}[htbp]
  \centering
  \includegraphics[width=.9\hsize]{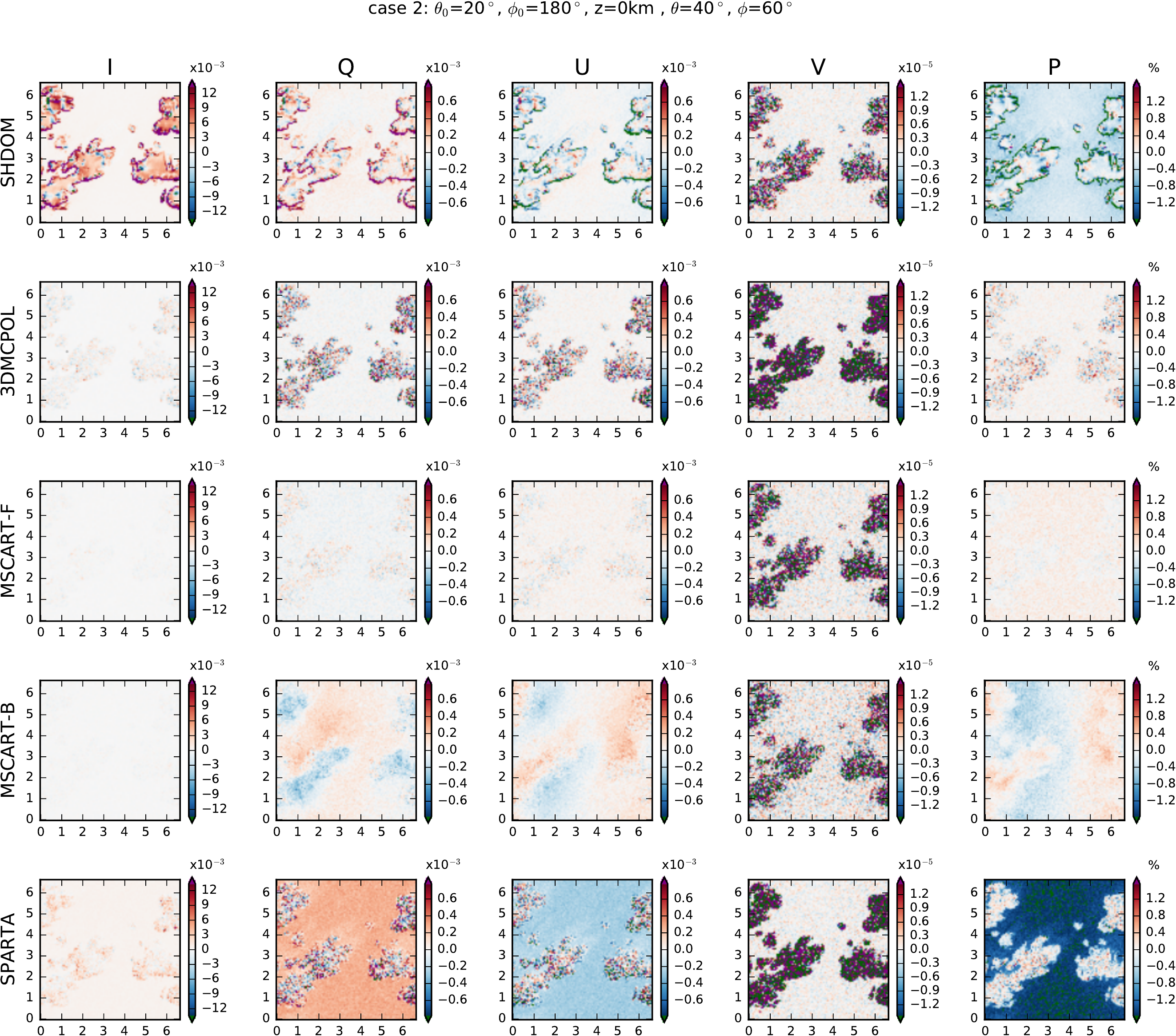}
  \caption{Absolute differences between individual model
    results and MYSTIC ($I_{i,model} - I_{i,MYSTIC}$) 
    for the cumulus cloud field in a molecular atmosphere with
    aerosols included (C3), case 2. Each row corresponds
    to a different model, see labels on the left. The labels on the x- and y-axes correspond to kilometers.
  }
  \label{fig:C3_diff_subcase2}
\end{figure*}
In order to investigate the reasons for differences revealed in the
statistics, we show the
absolute differences between MYSTIC and individual model results for
case 2 in Figure~\ref{fig:C3_diff_subcase2}.
The limits of the colorbar are set to 5\% of
the maximum value of the MYSTIC results for $I$ and $P$ and to 10\%
of the MYSTIC result for $|Q|$ and $|U|$. Values larger than the upper
limit are marked in purple and values smaller than the lower limit
are marked in green. 

SHDOM shows the largest differences at the cloud
boundaries as expected. But also the background shows systematic
differences: the degree of polarization obtained with SHDOM is
smaller than MYSTIC. 

For 3DMCPOL the difference plots show only
statistical noise, which is of course larger in the cloud regions than
in clearsky regions. 

Also MSCART-F agrees perfectly to MYSTIC.
Compared to 3DMCPOL, we find much less noise because
MSCART-F applies variance reduction methods. For MSCART-B we find
systematic differences for $Q$ and $U$ (inside the cloud regions
MYSTIC results are larger and outside the clouds MSCART-B results are
larger). These differences can again be attributed to the variance
reduction method of MSCART, which introduces a bias in backward
tracing mode. 

For SPARTA significant differences are found in particular in the
clear-sky region for $Q$ and $U$. The reason for these differences
could not be explained so far. It should be noted again, that the
differences occur only for a few specific geometries.   

In summary, we find a good agreement between the Monte Carlo models for
the cumulus cloud scenario. Obvious differences for a few of the cases
have been found between SPARTA and the other models, those should be
further investigated. SHDOM radiances are somewhat biased relative to those of
the Monte Carlo codes due to the lower angular resolution, and the
differences are larger on the cloud boundaries due to interpolating the
optical properties between grid points instead of assuming uniform cells
as for the Monte Carlo codes.

\section{Summary and Outlook}
\label{sec:summary}

We have presented the IPRT radiative transfer model intercomparison
for polarized radiative transfer in 3D geometry. Five models
participated; four of them (3DMCPOL, MSCART, MYSTIC, and SPARTA)
apply the Monte Carlo method to solve the
vector radiative transfer equation, one uses the spherical harmonics
discrete ordinate method (SHDOM).  

The first series of test cases are based on a simple two dimensional step cloud with
an optical thickness of 2 in one half and optical thickness of 18 in
the other half. We
simulated various sun-observer geometries, with sensor positions below
the cloud and above the cloud. The
results of this test demonstrate, that the polarized radiance is
significantly influenced by horizontal photon transport. 
Generally we find a very good agreement between the models within the
expected accuracy determined by the standard deviation of the Monte
Carlo results. We compared results obtained with the same number of
photons and tested the performance of variance reduction techniques as
included in MYSTIC and MSCART. For down-looking directions the
standard deviation is similar in the two models, for up-looking
directions, the MYSTIC variance reduction method VROOM is less efficient.
Apart from two exceptions for
specific geometries, where the model SHDOM shows significant
deviations, the models agree perfectly within the statistical noise.
As reference, MYSTIC has been
run with 10 times more photons. The relative standard deviation of the
reference simulation is below 0.2\% for $I$, below 1\% for $Q$ and
$U$ for down-looking direction, and below 5\% for $Q$ and
$U$ for up-looking directions. 

The second series of test cases are based on a single cubic cloud with an optical
thickness of 10. The first set of
simulations are for the cubic cloud in vacuum. Of particular interest
are simulations in the solar principal plane, for which the Stokes
vector components $U$ and $V$ are exactly 0 in 1D plane-parallel
geometry. The limited extension of the cloud breaks the symmetry between left side and
right side of the solar principal plane and characteristic
polarization patterns are generated. Furthermore it is interesting to see,
how the polarization changes with the sun-observer geometry. 
In the second set of simulations, the cubic cloud is surrounded by a
Rayleigh scattering layer with an optical thickness of 0.5. Here the
results show, how the polarization caused by cloud scattering
propagates in the clear-sky (Rayleigh scattering) part. Both cases
show that plane-parallel radiative transfer, which would result in
only two values, one for the cloud and one for the surrounding, can
only be a rough approximation. 
We have found a good quantitative agreement between the Monte Carlo
codes. MSCART agrees well to the other codes in forward tracing mode
but shows a significant bias in backward tracing mode, which can most
probably be attributed to the scattering-order dependent truncation method used
by MSCART. SPARTA deviates significantly for a few geometries, the
reason for this has not yet been found. 
SHDOM shows overall the same patterns as the Monte Carlo
models, but the \rms\ are larger than for the Monte Carlo codes
mainly due to the inherently lower angular and spatial resolution of a deterministic
representation code.
We again use the MYSTIC results as reference because they are accurate
and the variance reduction method is unbiased. The achieved accuracy
(relative standard deviation of MYSTIC results)
is of the order of 0.1\% for the unpolarized radiance $I$ and of the
order of 1\% for polarized radiance $Q$ and $U$. For $V$, the accuracy
is much less (more than 10\%) but still the characteristic patterns
could be compared qualitatively for models with variance reduction and
SHDOM and they were found to look identical.

The third series of test cases, the most realistic scenarios, are
based on a LES
cloud field which is surrounded by a standard molecular
atmosphere. A first set of simulations is without aerosols, in a second
set a realistic aerosol optical thickness profile is added. Due to
computational time limitations, the simulations were performed with 10
times less photons per pixel compared to the cubic cloud cases. We
found generally a good agreement between the Monte Carlo codes. 
MSCART-B polarized radiance results exibit a bias also for the
LES cloud field, which is smaller than in the cubic cloud
case. SHDOM shows the same patterns as the Monte Carlo codes, but the
quantitative comparison reveals differences particularly on cloud and
cloud shadow boundaries. As for the cubic cloud case,
SPARTA deviates from the other codes for a few specific cases
including the LES clouds.
The MYSTIC results, which were taken as reference, achieve an
accuracy (relative standard deviation)
of the order of 0.5\%-1\% for the unpolarized radiance $I$
and of the order of 5\% for the polarized radiance $Q$ and $U$. For $V$, the
relative standard deviation is in the range from 50\%--100\% for the
cases with aerosols, without aerosols even larger. Still, the
characteristic patterns are clearly visible and can be used for
qualitative comparisons with other models.

We have
not compared the computational times, because all simulations were
performed on different computation clusters, thus we can not decide, which of the
methods is the most efficient one. The computational time of a Monte
Carlo simulation depends on the required accuracy, because the standard
deviation of the result is proportional to $1/\sqrt{N_{\rm ph}}$. It
also depends on the spatial resolution:
in order to simulate an image of e.g. 100$\times$100
pixels we need 10000 times more photons than to simulate the domain
average radiance for the same scene for a given accuracy.
Furthermore, the CPU time depends
on the average optical thickness of the scene, which determines the
amount of multiple scattering. Roughly estimated, 
the computational times to reach a
relative standard deviation of about 1\% or less
for the Stokes components $I$, $Q$, and
$U$ vary from about 10 seconds per
pixel for the cubic cloud case 
to about 5 minutes per pixel for the step cloud and the cumulus
cloud scene. These times refer to MYSTIC simulations with variance
reduction on one processor (Intel(R) Xeon(R) CPU E5-2630 v4 @
2.20GHz). 

The results of all models are available on the IPRT webpage
(\url{http://www.meteo.physik.uni-muenchen.de/~iprt}) and the
supplementary material to this publication includes MYSTIC results of cases.
The webpage also provides
detailed descriptions of all test cases including the required input
data. 
 
IPRT has planned two further intercomparison projects: one
will focus on polarized radiative transfer in spherical geometry, the
other one on multiple-scattering lidar simulations.

\section*{Acknowledgment}
We thank the I3RC (Intercomparison of 3D Radiation Codes,
\url{https://i3rc.gsfc.nasa.gov}) team for
making available the input data of their intercomparison projects. The
step cloud and LES cloud field scenarios have been adapted from the
I3RC intercomparison project. 
 
\appendix

\section{Plots of the results of all testcases including cubic cloud (C2)
  and cumulus clouds (C3)}
\label{sec:app}

\begin{figure*}
  \centering
  \includegraphics[width=0.9\hsize]{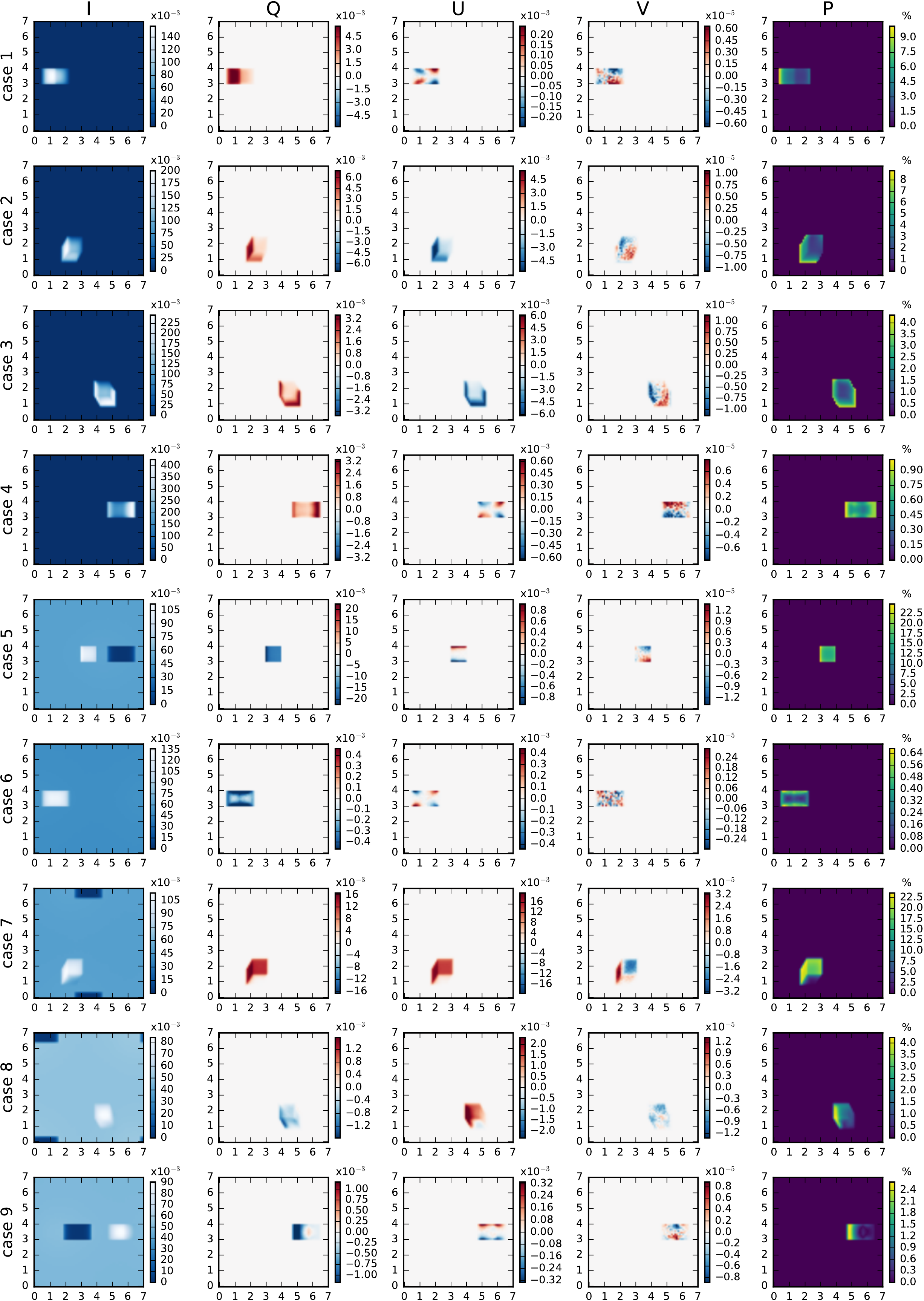}
  \caption{Results for scenario C2 (cubic cloud in vacuumn). }
  \label{fig:C2_all_noatm}
\end{figure*}

\begin{figure*}
  \centering
  \includegraphics[width=0.9\hsize]{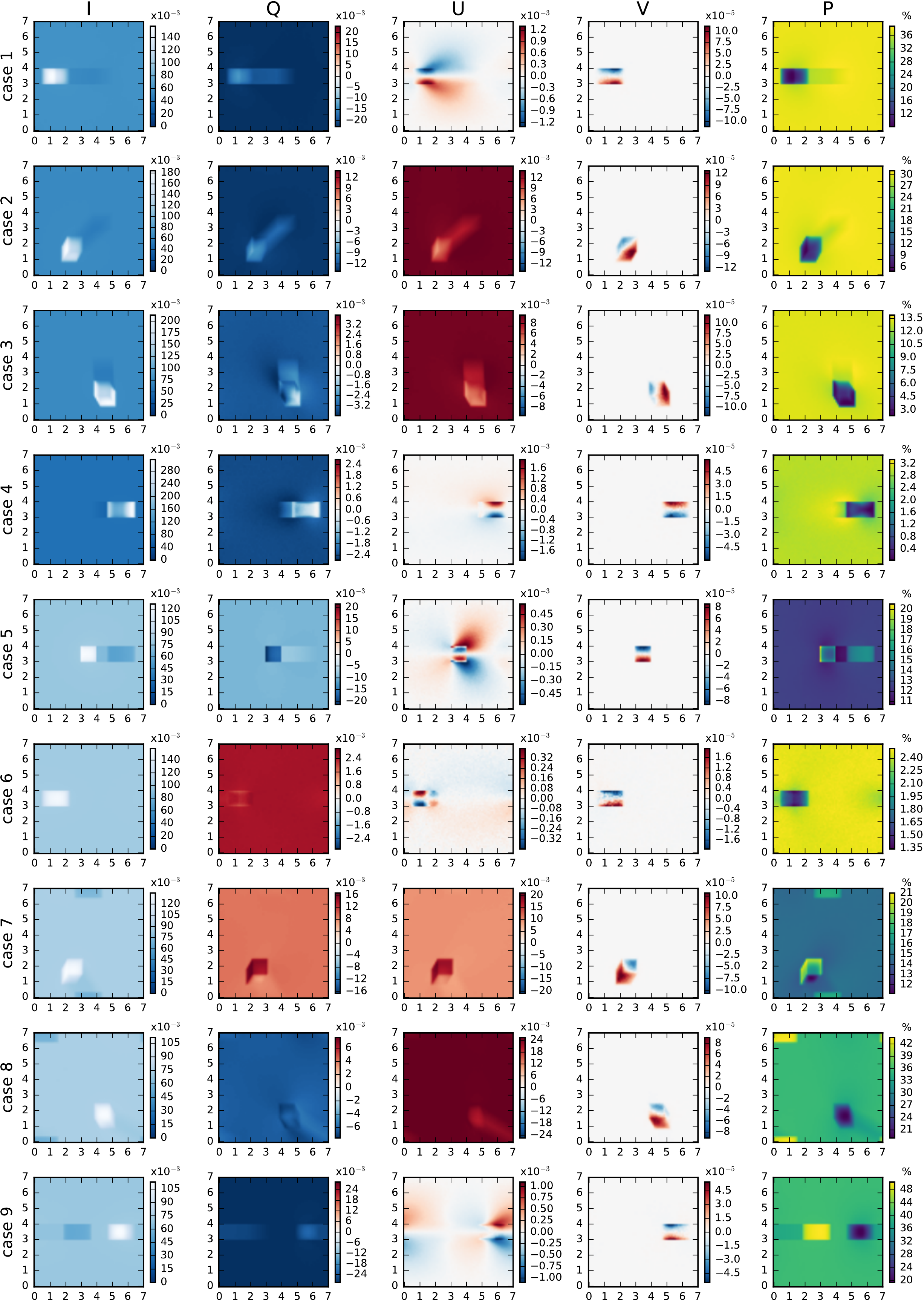}
  \caption{Results for scenario C2 (cubic cloud in Rayleigh scattering
    layer). }
  \label{fig:C2_all_atm}
\end{figure*}

\begin{figure*}
  \centering
  \includegraphics[width=0.9\hsize]{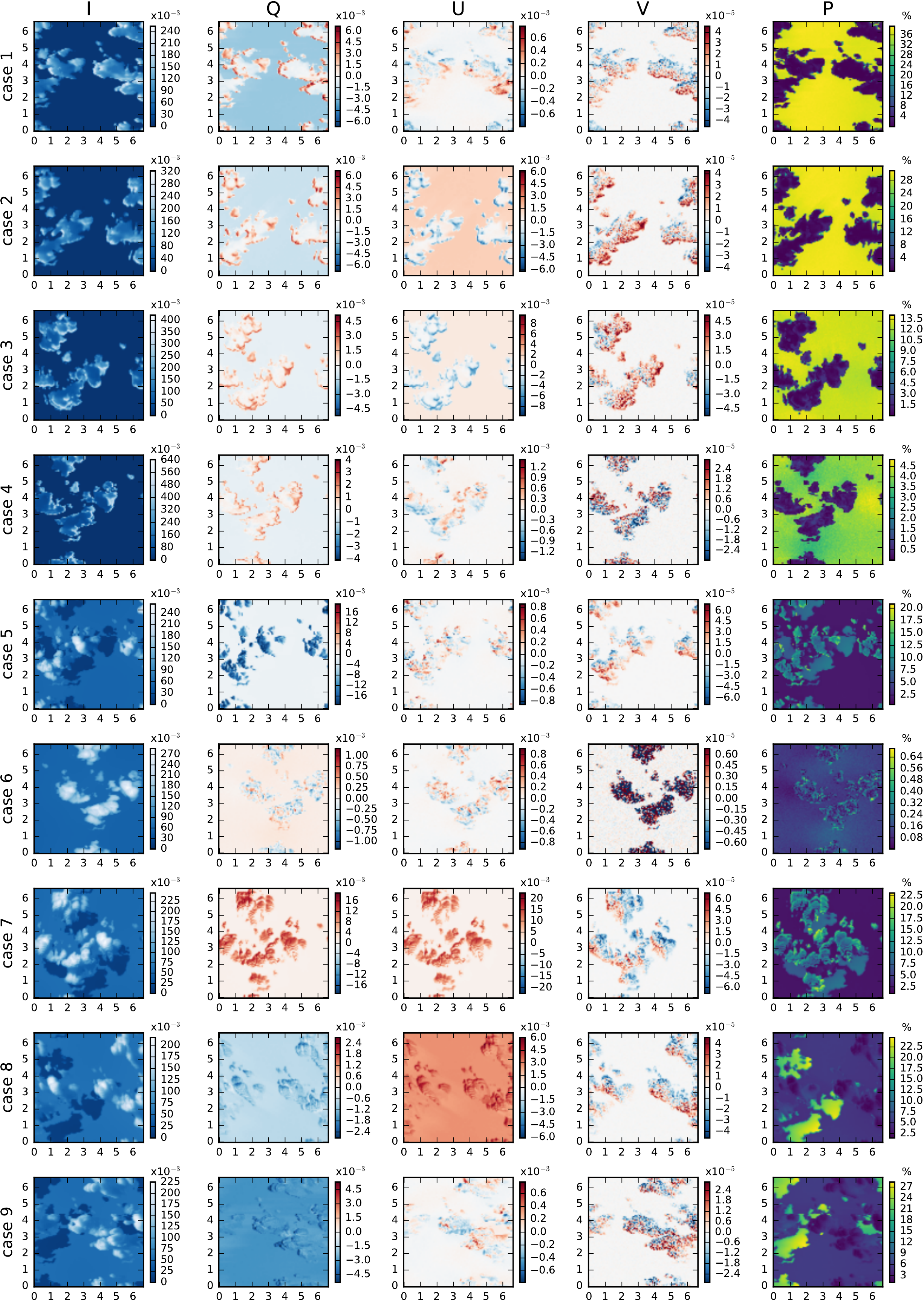}
  \caption{Results for scenario C3 (cumulus clouds surrounded by
    molecular atmosphere). }
  \label{fig:C2_all_noaer}
\end{figure*}

\begin{figure*}
  \centering
  \includegraphics[width=0.9\hsize]{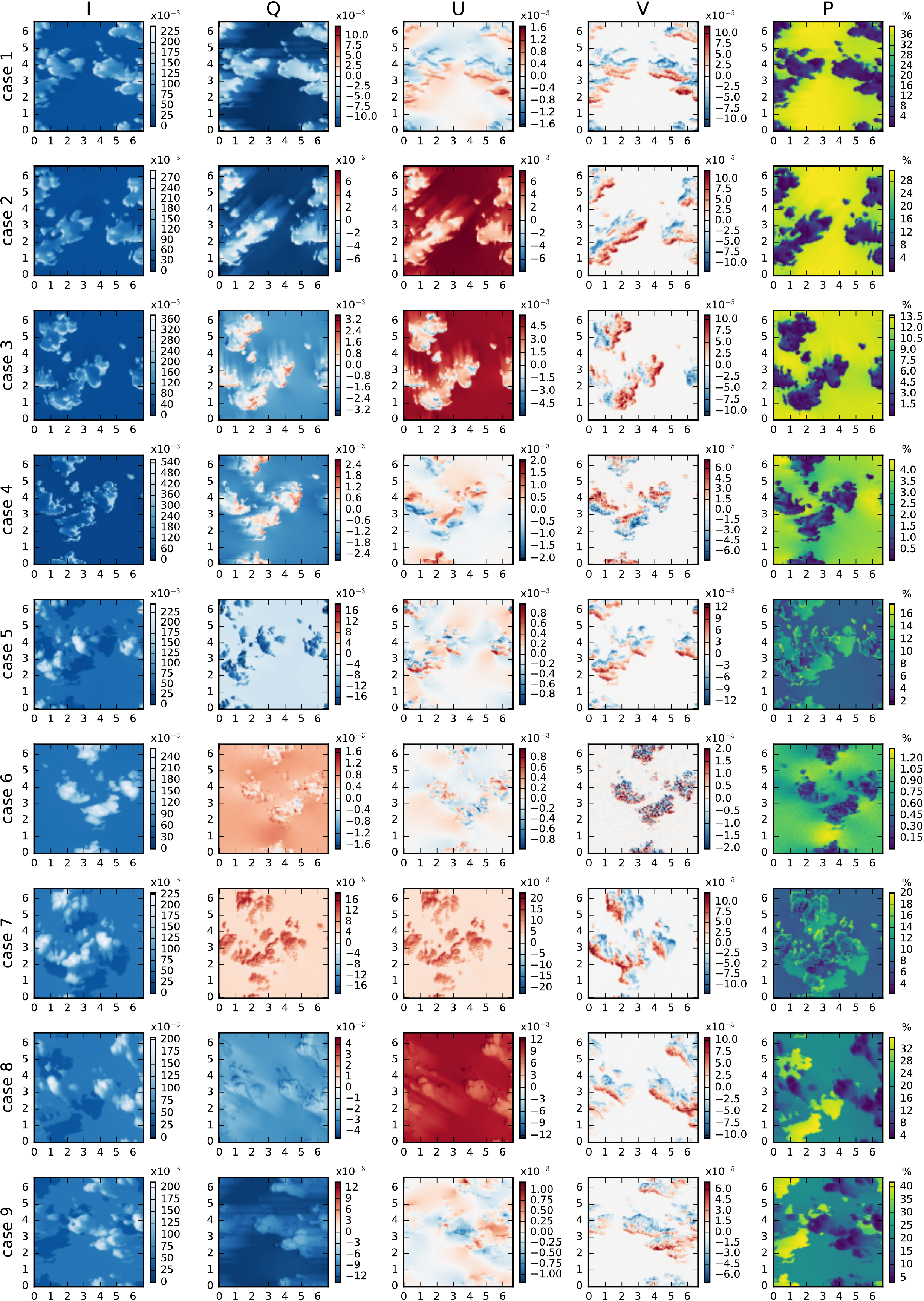}
  \caption{Results for scenario C3 (cumulus clouds in molecular
    atmosphere with aerosol).}
  \label{fig:C2_all_aer}
\end{figure*}

\bibliographystyle{elsarticle-num-names}
\bibliography{./literature.bib}

\end{document}